\documentclass[twoside,11pt]{article}
\usepackage{jair, theapa, rawfonts}

\ShortHeadings{Detecting Strategic Manipulation in Distributed Optimisation}
{Perez-Diaz, Gerding \& McGroarty}

\usepackage{amssymb,amsmath}
\usepackage{graphicx}
\DeclareGraphicsExtensions{.pdf,.png,.jpeg,.jpg,eps}
\graphicspath{{}}
\DeclareMathOperator*{\argmin}{arg\,min}

\usepackage{mathtools}
\DeclarePairedDelimiter\norm{\lVert}{\rVert}%
\usepackage[boxed]{algorithm2e}
\usepackage{placeins}
\usepackage{multirow}
\usepackage[implicit=false]{hyperref}
\usepackage{stackengine}

\DeclarePairedDelimiter\floor{\lfloor}{\rfloor}

\begin{document}
\title{Catching Cheats: Detecting Strategic Manipulation in Distributed Optimisation of Electric Vehicle Aggregators}
%\thanks{Supported by organization x.}}
%
\author{\name Alvaro Perez-Diaz \email a.perez-diaz@soton.ac.uk \\
\name Enrico Gerding \email eg@soton.ac.uk \\
\addr Electronics and Computer Science\\
University of Southampton
\AND
\name Frank McGroarty \email f.j.mcgroarty@soton.ac.uk \\
\addr Southampton Business School\\
University of Southampton}

\maketitle              % typeset the header of the contribution

\begin{abstract}
Given the rapid rise of electric vehicles (EVs) worldwide, and the ambitious targets set for the near future, the management of large EV fleets must be seen as a priority. Specifically, we study a scenario where EV charging is managed through self-interested EV aggregators who compete in the day-ahead market in order to purchase the electricity needed to meet their clients' requirements. With the aim of reducing electricity costs and lowering the impact on electricity markets, a centralised bidding coordination framework has been proposed in the literature employing a coordinator. In order to improve privacy and limit the need for the coordinator, we propose a reformulation of the coordination framework as a decentralised algorithm, employing the Alternating Direction Method of Multipliers (ADMM). However, given the self-interested nature of the aggregators, they can deviate from the algorithm in order to reduce their energy costs. Hence, we study the strategic manipulation of the ADMM algorithm and, in doing so, describe and analyse different possible attack vectors and propose a mathematical framework to quantify and detect manipulation. Importantly, this detection framework is not limited the considered EV scenario and can be applied to general ADMM algorithms. Finally, we test the proposed decentralised coordination and manipulation detection algorithms in realistic scenarios using real market and driver data from Spain. Our empirical results show that the decentralised algorithm's convergence to the optimal solution can be effectively disrupted by manipulative attacks achieving convergence to a different non-optimal solution which benefits the attacker. With respect to the detection algorithm, results indicate that it achieves very high accuracies and significantly outperforms a naive benchmark.
\end{abstract}

%%%%%%%%%%%%%%%%%%%%%%%%%%%%%%%%%%%%%%%%%%%%%%%%%%%%%
%%%%%%%%%%%%%%%%%%%%%%%%%%%%%%%%%%%%%%%%%%%%%%%%%%%%%
\section{Introduction}
\label{sec:intro}

To date, there exists a world-wide fleet of more than two million electric vehicles (EVs), combining purely electrical and hybrid \cite{iea2017}. Furthermore, EV sales are growing exponentially in most countries and there are targets to achieve 50 to 200 million of EVs at a global scale in the next decade \cite{iea2016}. These high penetration targets aim to reduce the use of fossil fuels and improve environmental conditions. However, the transition from conventional to electric vehicles is not without challenges \cite{Rigas2015}. Specifically, compared to traditional fuel powered vehicles, EVs present a novel and heavy strain to existing electricity networks, which will need to accommodate a new type of consumer with high demand.

In order to deal with this challenge, the last decade has seen the introduction of the concept of the EV aggregator \cite{Kempton2001}: an intermediary between a fleet of EVs and the electricity grid and markets. The aggregator is able to control the charging of its fleet, and this way informed collective decisions can be made. In contrast with individual EV operation, the much higher degree of coordination possible when a fleet is centrally managed by an aggregator offers great benefits. For example, electricity consumption to charge the fleet's batteries can be spread over time, avoiding expensive and polluting demand peaks. In particular, in this paper we focus on EV aggregators participating in the day-ahead market in order to purchase the electricity needed to meet their clients' energy requirements. In more detail, day-ahead markets match electricity supply and demand on an hourly basis (see Section \ref{sec:market}), and are the main source of wholesale electricity. Here, increased electricity demand means increased prices, resulting in the so-called \emph{price impact}, and hence it is in every market participant's interest to avoid unnecessary demand peaks.

In this work, we focus on a scenario where different EV aggregators co-exist in the same day-ahead market. These aggregators may vary in nature and size, but it is reasonable to assume that they are self-interested. Indeed, reduced electricity costs translate into more profit for the aggregator and/or more benefits for their EV fleet. In this scenario, reduced overall costs can be achieved by inter-aggregator coordination, producing more informed and optimised bidding. This coordination problem lies naturally under the umbrella of the multi-agent systems field \cite{Wooldridge2009}, and has been studied in the literature under a centralised approach by employing a centralised coordinator \cite{Perez-Diaz2018a,Perez-Diaz2018b}. More specifically, the focus in these works is to study payment mechanisms that incentivise truthful cooperation, using mechanism design and cooperative game theory, respectively. However, the proposed centralised approach requires a trusted environment where the participating aggregators report their private information to the central coordinator. In a realistic scenario, self-interested aggregators would be reluctant to share their private business information, thus presenting an important drawback to the proposed centralised approach.

In order to address this shortcoming, we propose a novel decentralised mechanism which allows the coordination of the EV aggregators without the need of a trusted coordinator, and without revealing their private requirement information. Specifically, we reformulate the centralised optimisation algorithm proposed by \citeauthor{Perez-Diaz2018a} \citeyear{Perez-Diaz2018a} using the Alternating Direction Method of Multipliers (ADMM), which decomposes the optimisation problem into smaller problems coordinated through an aggregation step \cite{Boyd2010}. Moreover, in order to provide transparency and remove the need for a trusted environment, the proposed algorithm can be implemented in a blockchain using smart contracts in a very similar vein as the work by \citeauthor{Munsing2017} \citeyear{Munsing2017}.

Although our proposed decentralised algorithm tackles the shortcomings described above, it introduces a new challenge. Specifically, in the decentralised case, the agents directly impact the computation of the optimal energy allocation. This introduces the possibility of strategic manipulation, where an aggregator deviates from the \emph{vanilla} ADMM algorithm with the aim of decreasing its energy costs, in detriment of the other aggregators. We explore this issue by defining several attack vectors which seek to improve an aggregator's own energy allocation. Furthermore, in order to address this problem, we propose a manipulation detection algorithm that monitors the behaviour of the aggregators to identify deviations. Note that this issue exists in any ADMM decentralised optimisation scenario with rational and self-interested agents, and is not limited to EV or smart grid studies. Also, we note that the manipulative issues described above are related to the area of distributed mechanism design \cite{Feigenbaum2002} where the aim is to design strategy-proof mechanisms (i.e. where agents cannot benefit by strategic manipulation) but where the computation is distributed. In such a setting, not only truthful preference elicitation but also faithful computation are required in order to obtain strategy-proof cooperation, and we focus on the latter in this work. More specifically, the proposed manipulation detection algorithm is naturally aligned with \emph{catch and punish} techniques in decentralised mechanism design, as described by \citeauthor{Shneidman2003} \citeyear{Shneidman2003}, which to date have not been applied to decentralised ADMM algorithms.

In more detail, this paper makes the following contributions to the state of the art:
\begin{itemize}
\item We propose the first decentralised optimisation algorithm for the coordination of self-interested EV aggregator participation in day-ahead markets.\footnote{Note that a preliminary study of this contribution was presented in the OptMAS workshop \cite{Perez-Diaz2018d}.}
\item We present the first study of strategic manipulation of the ADMM algorithm, where a self-interested agent tries to modify the algorithm's outcome in order to increase its personal utility.
\item We propose a detection algorithm to monitor the participating agents and find deviations form the vanilla ADMM algorithm.
\item We present a realistic case study to empirically evaluate both the decentralised coordination algorithm, the attack vectors and the detection algorithm. Results show that some of the considered attacks are able to reduce the attacker's energy costs by making the algorithm converge to sub-optimal allocations. Moreover, the proposed detection algorithm presents very good detection accuracy, up to and very close to 1, for scenarios with aggregators of the same size. However, considering mixtures of aggregators of different sizes is more challenging but the proposed algorithm is able to significantly outperform a naive benchmark in the vast majority of cases.
\end{itemize}

The rest of the paper is structured as follows. Section \ref{sec:lit} presents a literature review. Section \ref{sec:market} introduces the considered day-ahead market and the mathematical formalism to quantify price impact. Section \ref{sec:agg} details the considered EV aggregators and presents the proposed decentralised optimisation algorithm using ADMM. Next, a strategic manipulation study of the proposed ADMM algorithm is detailed in Section \ref{sec:attack}. Section \ref{sec:detectManipulation} presents the proposed mathematical formalism to detect strategic manipulation of the ADMM algorithm. Next, an empirical evaluation of the proposed algorithms using real market and driver data is detailed in Section \ref{sec:experiments}. Finally, we conclude in Section \ref{sec:conclusion}.

%%%%%%%%%%%%%%%%%%%%%%%%%%%%%%%%%%%%%%%%%%%%%%%%%%%%%
%%%%%%%%%%%%%%%%%%%%%%%%%%%%%%%%%%%%%%%%%%%%%%%%%%%%%
\section{Literature Review}
\label{sec:lit}

This paper builds upon existing literature in different fields, as detailed in this section. First, we consider works considering scenarios with several interacting EV aggregators. Secondly, we focus on the application of decentralised management techniques in smart grid scenarios. Thirdly, we consider the use of artificial intelligence techniques for smart EV charging. Fourth and lastly, we consider recent research on the manipulation of ADMM algorithms.

%%%%%%%%%%%%%%%%%%%%%%%%%%%%%%%%%%%%%%%%%%%%%%%%%%%%%
\subsection{Multi-Aggregator Scenarios}
\label{sec:multiAgg}

 A small body of literature addresses related multi-EV aggregator scenarios, as described below. \citeauthor{Qi2014} \citeyear{Qi2014} and  \citeauthor{Shao2016} \citeyear{Shao2016} study the hierarchical control of EV fleets where different aggregators are coordinated by a high-level coordinator. However, in their models the aggregators are not self-interested, and instead the authors focus on accommodating grid constraints and ensuring driver satisfaction. More related to our considered scenario, \citeauthor{Yu2016} \citeyear{Yu2016} study a setup where a number of EV aggregators can trade energy among them in order to fix forecasting deviations, instead of purchasing the energy from the grid. Although this is shown to improve the aggregators' energy costs, the authors do not consider price impact in their model and each aggregator performs independently. Another related study can be found in the work by   \citeauthor{Mukherjee2017} \citeyear{Mukherjee2017}. Their work considers a scenario where several private aggregators are present in a given city, and negotiate with each other in order to balance charging in the different limitedly available charging stations. The aim is to maximise the total number of EVs charged and the profit of the EV aggregators and their results indicate that coordinated operation improves the profit of the EV aggregators and the services offered to the drivers. Moreover, in a similar vein to our work,  \citeauthor{Wu2016} \citeyear{Wu2016} study a multi-aggregator day-ahead bidding scenario and apply game theory to find Nash equilibria. In more detail, each aggregator tries to categorise the other aggregators and thus forecast their day-ahead bids, and adjust their bidding accordingly. However, after introducing several approximations in order to simplify the model's structure, the proposed model depends on a complicated optimisation algorithm and does not guarantee existence of Nash equilibria. Finally, the same scenario considered in this paper is studied from a centralised perspective in the work by \citeauthor{Perez-Diaz2018a,Perez-Diaz2018b} \citeyear{Perez-Diaz2018a,Perez-Diaz2018b}. In these works, a number of self-interested aggregators perform coordinated bidding under the control of a centralised coordinator, and different payment mechanisms which incentivise cooperation rather than strategic manipulation are studied, using mechanism design and cooperative game theory, respectively. However, as discussed in Section \ref{sec:intro}, these centralised approaches require the aggregators to report all their private information to the coordinator, data that private entities would be reluctant to provide. In order to tackle this issue, the decentralised approach proposed in this paper removes the need for full information sharing, allowing coordination by revealing much less private information.

%%%%%%%%%%%%%%%%%%%%%%%%%%%%%%%%%%%%%%%%%%%%%%%%%%%%%
\subsection{Decentralised Management in the Smart Grid}
\label{sec:decSmartGrid}

Decentralised optimisation techniques have been widely applied in smart grid and power systems scenarios. In more detail, there is a body of literature studying decentralised charging scheduling of EVs  \cite{Ardakanian2014,Wen2012,Gan2013,Ma2013,LeFloch2015,LeFloch2016}. Overall, these works consider the problem of scheduling the charging of EVs in different decentralised fashions, considering each EV as an individual node in their respectively proposed algorithms. In more detail, \citeauthor{Ardakanian2014} \citeyear{Ardakanian2014} focuses on physical grid constraints, considering an electricity network managed by different access points, and its interaction with a fleet of EVs. Similarly, \citeauthor{Gan2013} \citeyear{Gan2013} and \citeauthor{Ma2013} \citeyear{Ma2013} consider decentralised valley-filling algorithms, where the aim is to flatten demand over time, and each EV sequentially updates its own charging schedule by iterative interaction with a central utility company. In a related vein, \citeauthor{Wen2012} \citeyear{Wen2012} consider a decentralised algorithm which employs discrete time intervals and selects subsets of EVs to be charged at each time interval, by iterative communication between each EV and their aggregator. Finally, \citeauthor{LeFloch2015,LeFloch2016} \citeyear{LeFloch2015,LeFloch2016} consider \emph{vehicle-to-grid} (V2G) scenarios, where the EVs are able to inject energy back to the grid when needed. Although all these works study different aspects of EV charging scheduling under decentralised algorithms, they do not consider the interaction among different self-interested aggregators, which is one of the aims of this paper.

Furthermore, decentralised algorithms have been employed in many non-EV related smart grid publications. As discussed in Section \ref{sec:intro}, an algorithm that has acquired great popularity in recent years due to its versatility and good convergence properties is ADMM \cite{Boyd2010}. It has been employed in multitude of smart grid studies, such as power flow \cite{Wang2017,Sulc2014,Peng2014,Scott2014}, demand response \cite{Nguyen2018,Zhou2016} and micro-grid \cite{Munsing2017} scenarios. However, this algorithm (or variants) has not been employed to study the decentralised coordination of self-interested aggregators.

%%%%%%%%%%%%%%%%%%%%%%%%%%%%%%%%%%%%%%%%%%%%%%%%%%%%%
\subsection{Artificial Intelligence and EV Charging}
\label{sec:AI_lit}

The field of artificial intelligence has also devoted a substantial amount of effort to solve EV-related issues. Examples include smart charging scheduling algorithms \cite{DeWeerdt2018,Gerding2019}, preference elicitation from drivers \cite{Gerding2011,Stein2012,Robu2013,Hayakawa2015,Gerding2016} and centralised coordination of EV aggregators \cite{Perez-Diaz2018a,Perez-Diaz2018b}. In more detail, the first area is concerned with the problem of allocating or scheduling the available electricity to a fleet of EVs. This is a complex problem that takes into account each EV's energy requirements and time constraints, and has been studied from both offline and online perspectives. Furthermore, the second area of research considers the EV charging problem as multi-agent system where the drivers are self-interested and the EV aggregators elicit their driving needs. These include the amount of electricity needed and time constraints. These works use techniques from mechanism design in order to truthfully elicit these preferences, both in offline and online settings, and in a variety of scenarios. In contrast to these two fields of research, this work deals with a previous step in the charging problem, the purchasing of electricity by the EV aggregators, who can afterwards use scheduling algorithms to allocate their obtained energy and use preference elicitation models to encourage drivers to truthfully reveal their needs. Finally, the third body of literature considers the same scenario as this current work but from a centralised perspective, and has been already detailed in Section \ref{sec:multiAgg}.

%%%%%%%%%%%%%%%%%%%%%%%%%%%%%%%%%%%%%%%%%%%%%%%%%%%%%
\subsection{Manipulation of ADMM Algorithms}
\label{sec:manipADMM}

As described in Section \ref{sec:decSmartGrid}, decentralised optimisation algorithms are widely used, not only in smart grid related studies, but also in most technical fields. The reasons range from better scaling to large problem sizes to preserving privacy. There is, however, a gap between the introduction of such algorithms and the study of their robustness to potential manipulative/malicious attacks \cite{Munsing2018}. Specifically, in contrast with centralised algorithms, in the decentralised case, each node or agent participating in the algorithm will perform part of the calculations, or will transmit messages to a coordinator, hence the possibilities of cyber-attack or manipulation increase. In order to address these important issues, a few works have been published in recent years studying related topics, which will be described next. Note that we focus on ADMM algorithms, but the findings of all these studies should be generalizable to other iterative decentralised optimisation methods.

Following the work by \citeauthor{Munsing2018} \citeyear{Munsing2018}, we can classify the existing literature based on the technique employed:
\begin{itemize}
\item \emph{Round-robin techniques} \cite{Liao2016,Liao2017}: these techniques seek to identify compromised nodes by replacing the coordination step of the ADMM algorithm by a round-robin detection algorithm which compares the proposals of different subsets of nodes in order to identify discrepancies. Once corrupted nodes have been identified, the coordinator switches back to the ADMM algorithm.
\item \emph{Filtering techniques} \cite{Liao2018}: these techniques do not try to identify compromised nodes, but to employ robust statistics and outlier detection techniques in order to accurately compute the desired global quantities even in the presence of malignant data.
\item \emph{Non-linear weighting techniques} \cite{Chen2018}: similarly to filtering techniques, these techniques also do not try to identify compromised nodes. Instead, they employ data from all nodes, but introduce weights to scale down the impact of suspicious nodes.
\item \emph{Convexity techniques} \cite{Munsing2018}: these techniques detects compromised nodes and false-data injection in convex algorithms by checking for convexity violations. 
\end{itemize}
Overall, these works focus on cyber-security, \emph{i.e.} the effects of external attacks which compromise an internal node (a participant in the ADMM algorithm). Moreover, the papers discussed above focus on random noise injection by a malignant agent, which prevents the algorithm from converging. In contrast, in this work, we study strategic manipulation of the ADMM algorithm by internal self-interested agents. In more detail, rather than considering external malignant attackers, we consider algorithm participants that want to achieve more beneficial outcomes for themselves, even if this is in detriment of the other participants, and deviate from the vanilla ADMM algorithm in order to do so. This differs from these existing works in two aspects: (i) the algorithm can still converge to a stable outcome, (ii) the manipulating agents will use clever cheating techniques, as injecting random noise will not be beneficial for them.

%%%%%%%%%%%%%%%%%%%%%%%%%%%%%%%%%%%%%%%%%%%%%%%%%%%%%%%%%%%%%%%%%%%%%%%
%%%%%%%%%%%%%%%%%%%%%%%%%%%%%%%%%%%%%%%%%%%%%%%%%%%%%%%%%%%%%%%%%%%%%%%
\section{The Day-Ahead Market}
\label{sec:market}

This section details the day-ahead market structure considered in this paper and present in most countries. Moreover, we discuss how to quantify the price impact of buy orders (electricity demand), which is an important aspect of our work. The exposition in this section follows the work by \citeauthor{Perez-Diaz2018a,Perez-Diaz2018b} \citeyear{Perez-Diaz2018a,Perez-Diaz2018b}.

\begin{figure}[!b]
\centering
        \stackunder[6pt]{\includegraphics[width=.32\linewidth]{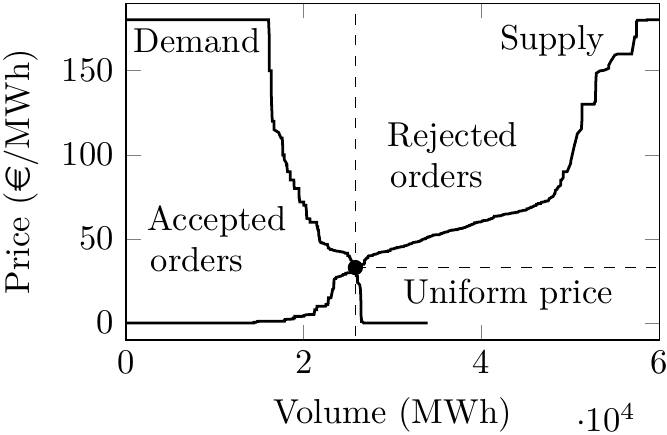}}{(a)}%
        \stackunder[6pt]{\includegraphics[width=.32\linewidth]{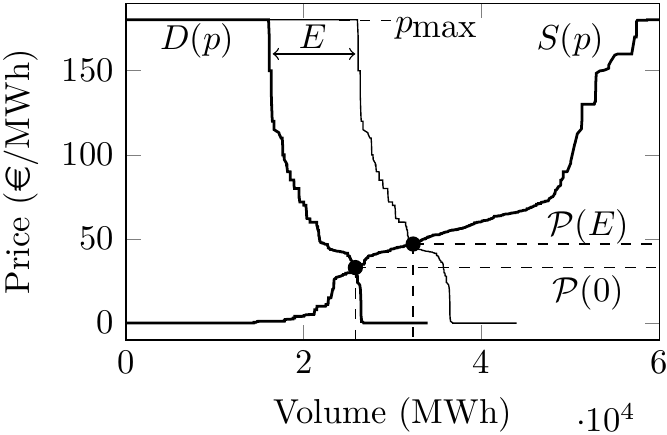}}{(b)}%
        \stackunder[6pt]{\includegraphics[width=.32\linewidth]{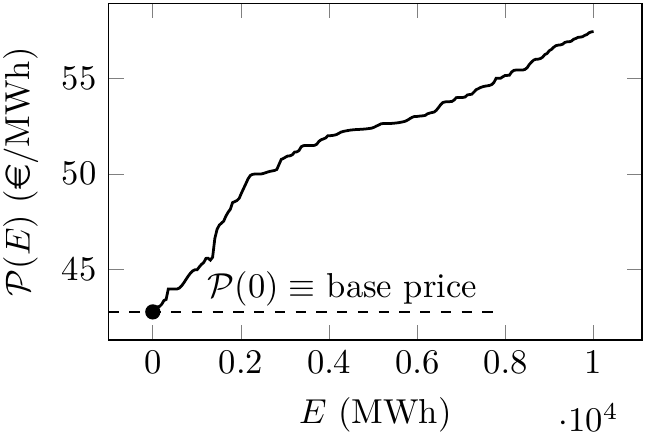}}{(c)}%
        \vspace{-0.3cm}
        \caption{(a) Aggregated supply and demand curves, and market clearing mechanism. (b) Price impact of a buy order with volume $E$ and maximum price $p_\text{max}$. (c) Final price function $\mathcal{P}(E)$. Source: OMIE, 01/11/2016, 11$^{\text{th}}$ hour.}
        \label{fig:market}
\end{figure}

Day-ahead markets divide each day into 24 hourly slots, each running a separate uniform-priced double-sided auction. Before closure time (usually noon) on a given day, bids and offers for each hourly slot of the next day must be submitted to the market. Then, a matching algorithm determines the accepted bids and offers, and establishes an hourly uniform price using marginal pricing, this is, the price of the intersection between supply and demand.

Bids (buy orders) and offers (sell orders) for each hourly slot are quantity-price pairs. For bids (offers), the price represents the highest (lowest) price the participant is willing to pay (sell for). As is common in most markets, we define a minimum price $p_\text{min} = 0$ and some maximum price, $p_\text{max}$. After closure time, the auctioneer aggregates all buy and sell orders, by high-price and low-price priorities, respectively. This generates the aggregated demand and supply curves, and their intersection determines the accepted orders and the resulting uniform price, as depicted in Fig. \ref{fig:market} (a).

Clearly, the arrival of a new buy order pushes the clearing price up if it gets accepted (\emph{i.e.} if it lies towards the left-hand side of the intersection). Fig. \ref{fig:market} (b) illustrates the effect of a new buy order with quantity $E$ placed at price $p_\text{max}$. The price increase (price impact) depends on the new order's price and quantity, and on the supply and demand curves. Price impact is an essential market characteristic associated with large market participants, and careful managing is required to avoid pushing prices up unnecessarily. Price impact has been studied in the electricity markets literature by employing residual curves \cite{Herranz2012,Perez-Diaz2018a,Perez-Diaz2018b,Perez-Diaz2018d}, which are detailed below.

Employing standard notation, for any given hour $t$, let $D_t(p)$ and $S_t(p)$ be the aggregated demand and supply curves respectively, as a function of price, $p$. The residual supply curve is defined as $R_t(p) = S_t(p)-D_t(p)=E$, and represents the amount of energy, $E$, an agent could bid for while maintaining a clearing price $p$. Conversely, the clearing price when bidding a quantity $E$ is given by $p = R_t^{-1}(E)$. Introducing the notation $\mathcal{P}_t(E) = R_t^{-1}(E)$, the clearing price when the new agent bids an amount $E$ is
$p = \mathcal{P}_t(E)$, and the price impact $\Delta p$ of this order is given by $\Delta p = \mathcal{P}_t(E) - \mathcal{P}_t(0)$,
where $\mathcal{P}_t(0)$ represents the \emph{base price} at hour $t$, i.e. the price without the agent's new bid. This formalism is depicted in Figs. \ref{fig:market} (b) and (c).

We are now ready to introduce the EV aggregator model considered in this paper and the optimal day-ahead bidding algorithm.

%%%%%%%%%%%%%%%%%%%%%%%%%%%%%%%%%%%%%%%%%%%%%%%%%%%%%%%%%%%%%%%%%%%%%%%
%%%%%%%%%%%%%%%%%%%%%%%%%%%%%%%%%%%%%%%%%%%%%%%%%%%%%%%%%%%%%%%%%%%%%%%
\section{Optimal Multi-EV Aggregator Participation in the Day-Ahead Market}
\label{sec:agg}

As discussed in Section \ref{sec:intro}, an EV aggregator is responsible for the charging of a fleet of EVs and, to this end, purchases the required electricity from the day-ahead market (see Section \ref{sec:market}). We will start by describing the considered aggregator structure and operation. Then, we will describe the optimal bidding algorithm proposed by \citeauthor{Perez-Diaz2018a,Perez-Diaz2018b,Perez-Diaz2018d} \citeyear{Perez-Diaz2018a,Perez-Diaz2018b,Perez-Diaz2018d} and how it can be used to optimise the bidding of a group of EV aggregators with a central coordinator. Finally, we will decompose this centralised algorithm into a decentralised optimisation algorithm by using the Alternating Direction Method of Multipliers (ADMM), as discussed in Section \ref{sec:intro}.

%%%%%%%%%%%%%%%%%%%%%%%%%%%%%%%%%%%%%%%%%%%%%%%%%%%%%%%%%%%%%%%%%%%%%%%%
\subsection{EV Aggregator Model}
\label{sec:aggModel}

In our model, following the work by \citeauthor{Bessa2012b} \citeyear{Bessa2012b} and \citeauthor{Perez-Diaz2018a,Perez-Diaz2018b,Perez-Diaz2018d} \citeyear{Perez-Diaz2018a,Perez-Diaz2018b,Perez-Diaz2018d}, EVs arrive and depart dynamically over time. When an EV $i$ arrives to the charging point, it communicates the desired departure time, $t_d^i$, and desired state of charge at departure, $\text{SoC}_d^i$, to the aggregator. We assume that arrival time and state of charge, $t_0^i$ and $\text{SoC}_0^i$ can be automatically inferred by the aggregator. Each EV has a maximum charging speed, $P_\text{max}^i$ in kW, which depends on two factors: the available physical infrastructure, and the EV's battery. The charging schedule of the EV is then left at the aggregator's discretion, which can choose when to perform the charging while guaranteeing the desired state of charge by departure time. This flexibility allows charging the battery in an informed way, rather than randomly, or at arrival, providing cheaper electricity costs.

Due to the nature of the day-ahead market, electricity bids need to be placed between 12 and 36 hours before delivery time (assuming market closure at noon, see Section \ref{sec:market}). This requires the market participants to forecast their electricity needs, as described next, and bid accordingly.

Again, following the work by  \citeauthor{Bessa2012b} \citeyear{Bessa2012b} and \citeauthor{Perez-Diaz2018a,Perez-Diaz2018b,Perez-Diaz2018d} \citeyear{Perez-Diaz2018a,Perez-Diaz2018b,Perez-Diaz2018d}, we model the requirements of an EV $i$ by employing two vectors with 24 entries each, $\mathbf{r}^{\text{min}, i}$ and $\mathbf{r}^{\text{max}, i}$. Specifically, $r_t^{\text{min}, i}$ is the amount of energy needed at hour $t$ assuming charging has been left for the last possible moment and that the charging requirements need to be fulfilled. Conversely, $r_t^{\text{max}, i}$ is the amount of energy needed at hour $t$ assuming charging starts as soon as possible. For example, consider an EV arriving at 3pm, stating 9pm departure time and 8kWh charging needs with $P_\text{max} = 3$kW. Then, $\mathbf{r}^{\text{min}, i}$ would be as specified in Table \ref{table:Rs}. Specifically, if 6pm is reached with no charging done, at least 2kW of energy needs to be charged between 6-7pm in order to fulfil the EV driver requirements. The same applies with 3kW between 7-8pm and 8-9pm. Similarly, for the same scenario, the requirement vector $\mathbf{r}^{\text{max}, i}$ would be as specified in Table \ref{table:Rs}.

Then, in order to provide mathematical tractability, two global energy requirement vectors, $\mathbf{R}^{\text{min}}$ and $\mathbf{R}^{\text{max}}$, can be obtained by summing the hourly requirements of all the EVs associated to the particular aggregator, \emph{i.e.} $R_t^{\text{min}} = \sum_{i=1}^N r^{\text{min}, i}_t$ and $R_t^{\text{max}} = \sum_{i=1}^N r^{\text{max}, i}_t$. Note that these aggregated constraints do not exactly capture the individual requirements of each EV, but have been widely employed in the literature \cite{Bessa2012b,Bessa2013a,Bessa2013b,GonzalezVaya2015,Perez-Diaz2018a,Perez-Diaz2018b,Perez-Diaz2018d}. The reasons are the fact that considering constraints for each individual EV renders the problem unfeasible with moderate problem sizes, and the fact that bidding uses day-ahead price and energy requirements forecasts, which will not be exact anyway.

We will denote the quantities that need to be forecasted with a hat: hourly energy requirements, $\hat{R}_t^{\text{min}}$ and $\hat{R}_t^{\text{max}}$, hourly number of available EVs, $\hat{N}_t$, and hourly price impact functions, $\hat{\mathcal{P}}_t$.

\begin{table}[!h]
\centering
\begin{tabular}{|c|c|c|c|c|c|c|}
\hline
$r^{\text{min}, i}_3$ & $r^{\text{min}, i}_4$ & $r^{\text{min}, i}_5$ & $r^{\text{min}, i}_6$ & $r^{\text{min}, i}_7$ & $r^{\text{min}, i}_8$ & $r^{\text{min}, i}_9$    \\ \hline
0     & 0     & 0     & 2     & 3     & 3     & 0    \\ \hline
\end{tabular}\\[0.1cm]
\begin{tabular}{|c|c|c|c|c|c|c|}
\hline
$r^{\text{max}, i}_3$ & $r^{\text{max}, i}_4$ & $r^{\text{max}, i}_5$ & $r^{\text{max}, i}_6$ & $r^{\text{max}, i}_7$ & $r^{\text{max}, i}_8$ & $r^{\text{max}, i}_9$    \\ \hline
3     & 3     & 2     & 0     & 0     & 0     & 0    \\ \hline
\end{tabular}
\caption{Example of requirement vectors  $\mathbf{r}^{\text{min}, i}$ and $\mathbf{r}^{\text{max}, i}$}
\label{table:Rs}
\end{table}

%%%%%%%%%%%%%%%%%%%%%%%%%%%%%%%%%%%%%%%%%%%%%%%%%%%%%%%%%%%%%%%%%%%%%%%%
\subsection{Optimal Day-Ahead Bidding Algorithm}
\label{sec:bidding}

Now that the day-ahead and EV aggregator models have been detailed, we are ready to present the optimal day-ahead bidding algorithm. The algorithm is from the work by \citeauthor{Perez-Diaz2018a,Perez-Diaz2018b,Perez-Diaz2018d} \citeyear{Perez-Diaz2018a,Perez-Diaz2018b,Perez-Diaz2018d} and reproduced here for convenience. The mathematical problem is defined as follows: given an EV aggregator's forecasted requirements and price impact functions, find the optimal distribution of energy quantities to bid across the 24 hourly slots of the next day, $\mathbf{E} = \left( E_0, \ldots, E_{23}\right)$, in order to satisfy its clients' charging needs while minimising the total cost of the purchased energy. We assume that the agent's bids are set at maximum price, $p_{\text{max}}$, in order to guarantee execution. Hence only bidding hours and quantities need to be decided.

As discussed in \cite{Perez-Diaz2018a}, and in order to avoid a complex minimisation landscape with multiple minima, the forecasted hourly price impact functions $\hat{\mathcal{P}}_t$ (see Sections \ref{sec:market} and \ref{sec:aggModel}) are approximated by quadratic convex functions. This makes the optimisation problem convex while barely affecting the accuracy of the algorithm \cite{Perez-Diaz2018a}. Specifically, the convex price impact functions are given by $\hat{\mathcal{P}}_t^{\text{convex}} = a_t E_t^2 + b_t E_t + \hat{\mathcal{P}}_t(0)$, where all the coefficients $a_t$ are restricted to be positive. Formally, the optimisation algorithm is given by Eqs.~(\ref{eq:PI}), (\ref{eq:PI1}), (\ref{eq:PI2}), (\ref{eq:PI3}). In more detail, the objective function (\ref{eq:PI}) minimizes the total cost of the purchased energy. The constraints guarantee that the amount of purchased energy is enough to satisfy the forecasted demand (\ref{eq:PI1}), that it is not purchased before the forecasted arrival of the EVs (\ref{eq:PI2}) and that the energy purchased at each hour is not greater than the amount that the aggregator is able to charge at the given hour, based on the forecasted number of available vehicles (the aggregator cannot store energy). It is worth noting that the number of constraints is always 72, independent on the fleet size. Also, given the convexity of the problem, there exists a unique global minimum, which we are guaranteed to find.
\vspace{-.25cm}
\begin{subequations}
\begin{align}
\min_{\{E_t\}} \sum_t \hat{\mathcal{P}}_t^{\text{convex}} (E_t) \cdot E_t \label{eq:PI}\\
\sum_{j=0}^t E_j \geq \sum_{j=0}^t \hat{R}_j^{\text{min}} \, , \,\, \forall t = 0,\ldots,23 \label{eq:PI1}\\
\sum_{j=0}^t E_j \leq \sum_{j=0}^t \hat{R}_j ^{\text{max}}\, , \,\, \forall t = 0,\ldots,23 \label{eq:PI2}\\
E_t / 1~\text{hour} \leq \hat{N}_t P_\text{max} \, , \,\,  \forall t = 0,\ldots,23 \label{eq:PI3}\\
E_t \geq 0 \, , \,\,  \forall t = 0,\ldots,23 \label{eq:PI4}
\end{align}
\end{subequations}

%%%%%%%%%%%%%%%%%%%%%%%%%%%%%%%%%%%%%%%%%%%%%%%%%%%%%%%%%%%%%%%%%%%%%%%%
\subsection{Centralised Joint Bidding}
\label{sec:jointBidding}

The bidding algorithm detailed in the previous section for a single aggregator can be extended to perform joint bidding, where a coordinator collects the requirements of a number of independent aggregators and applies the optimisation algorithm globally. In more detail, consider a set of $n$ EV aggregators. Then, following \cite{Perez-Diaz2018a,Perez-Diaz2018b,Perez-Diaz2018d} and overloading the variable $i$,  let $\hat{R}^{\text{min}, i}_t$ and $\hat{R}^{\text{max}, i}_t$ be aggregator $i$'s forecasted energy requirements for hour $t$, and $\hat{N}^ i_t$ the number of available EVs from aggregator $i$, as specified in Section \ref{sec:bidding}. The combined requirements of all the aggregators are then:

\begin{minipage}{.33\linewidth}
\begin{gather}
  \hat{R}^\text{min}_t = \sum_{i = 1}^n \hat{R}^{\text{min}, i}_t
  \label{eq:combRmin}
  \raisetag{10pt}
\end{gather}
\end{minipage}%
\begin{minipage}{.33\linewidth}
  \begin{gather}
  \hat{R}^\text{max}_t = \sum_{i = 1}^n \hat{R}^{\text{max}, i}_t
  \label{eq:combRmax}
  \raisetag{10pt}
  \end{gather}
\end{minipage}%
\begin{minipage}{.33\linewidth}
  \begin{gather}
  \hat{N}_t = \sum_{i  = 1}^n \hat{N}^ i_t \,\,\,\,\,\,\,\,\,\,\,
  \label{eq:combN}
  \raisetag{10pt}
  \end{gather}
\end{minipage}

To find the optimal global energy bids, the bidding optimisation algorithm given by Eqs.~(\ref{eq:PI}), (\ref{eq:PI1}), (\ref{eq:PI2}), (\ref{eq:PI3}), (\ref{eq:PI4}) can be applied with constraints given by the combined requirements (\ref{eq:combRmin}), (\ref{eq:combRmax}) and (\ref{eq:combN}). This will result in obtaining a global day-ahead energy volume $E_t$ for each hour $t$, which can be then distributed among the $n$ aggregators.

The redistribution mechanism is defined in \cite{Perez-Diaz2018a}, and allocates an hourly energy schedule to each participating aggregator after obtaining a global energy schedule as detailed above. The redistribution problem is as follows. Letting $E^i_t$ be the amount of energy allocated to EV aggregator $i$ at time $t$, we need to find $E^i_t$ for $t=0,\ldots, 23$ and $i=1,\ldots,n$ satisfying the following constraints:
\begin{subequations}
\begin{align}
& \sum_{j=0}^t E^i_j \geq \sum_{j=0}^t \hat{R}_j^{\text{min}, i}  , \, \forall t = 0,\dots,23  ; \, \forall i=1,\dots,n  \label{eq:dist1}\\
& \sum_{j=0}^t E^i_j \leq \sum_{j=0}^t \hat{R}_j ^{\text{max}, i}  , \, \forall t = 0,\dots,23  ; \, \forall i=1,\dots,n  \label{eq:dist2}\\
& E^i_t / 1~\text{hour} \leq \hat{N}^i_t P_\text{max} , \,\, \forall t = 0,\ldots,23 ; \, \forall i=1,\ldots,n  \label{eq:dist3}\\
& E^i_t  \geq 0 , \,\, \forall t = 0,\ldots,23 ; \, \forall i=1,\ldots,n \label{eq:dist4}\\
& \sum_{i=1}^n E^i_t  = E_t , \,\, \forall t = 0,\ldots,23 \label{eq:dist5}
\end{align}
\end{subequations}

In this constraint satisfaction problem, Eqs.~(\ref{eq:dist1}), (\ref{eq:dist2}), (\ref{eq:dist3}), (\ref{eq:dist4}) ensure that each EV aggregator has enough energy to satisfy its requirements, no more, no less, for each hour. Eq.~(\ref{eq:dist5}) makes sure the sums of the allocated hourly energies add up to the available global energy.

%%%%%%%%%%%%%%%%%%%%%%%%%%%%%%%%%%%%%%%%%%%%%%%%%%%%%%%%%%%%%%%%%%%%%%%%
\subsection{Decentralised Optimisation Algorithm}
\label{sec:decOpt}

We are now ready to introduce the novel decentralised optimisation algorithm based on ADMM \cite{Boyd2010}. Specifically, our goal is to reformulate the optimisation problems given by Eqs.~(\ref{eq:PI}), (\ref{eq:PI1}), (\ref{eq:PI2}), (\ref{eq:dist1}), (\ref{eq:dist2}), (\ref{eq:dist3}), (\ref{eq:dist4}), (\ref{eq:dist5}) as an iterative decentralised algorithm, where each EV aggregator solves a local optimisation problem using only their own private information. The solutions to each local problem are coordinated by a global \emph{consensus} step, and this procedure is iterated. \emph{Consensus} refers to the fact that, asymptotically, all the local variables will coincide. This type of algorithm is appropriate in our setting for several reasons: (i) given that our problem is convex, it is guaranteed to converge to the global optimum \cite{Boyd2010}; (ii) it enables coordination without the aggregators revealing their energy requirements, \emph{i.e.} $\mathbf{R}^{\text{min}, i}$ and $\mathbf{R}^{\text{max}, i}$; (iii) it is particularly well suited for blockchain implementation, providing transparency and anti-tampering guarantees \cite{Perez-Diaz2018d,Munsing2017}.

Following the notation introduced in Section \ref{sec:jointBidding}, recall that $\mathbf{E^i} = \left( E^i_0, \ldots, E^i_{23}\right)$ denotes the energy schedule for aggregator $i$. Moreover, let $\mathbf{E} = \left( \mathbf{E}^1, \ldots, \mathbf{E}^n \right)$ be the joint vector encapsulating each individual energy schedule. We can now rewrite Eq.~(\ref{eq:PI}) as:
\begin{equation}
\min_{\mathbf{E}} \sum_{t=0}^{23} \left[\hat{\mathcal{P}}_t^{\text{convex}} \left(\sum_{i=1}^n E^i_t \right) \cdot \sum_{i=1}^n E^i_t \right]
=  \min_{\mathbf{E}} \sum_{i=1}^{n} \left[  \sum_{t=0}^{23} \left( E_t^i \cdot \hat{\mathcal{P}}_t^{\text{convex}} \left( \sum_{j=1}^n E^j_t \right) \right) \right] \label{eq:OF_split}
\end{equation}
This way the objective function is expressed as a sum of $n$ terms, as required by the ADMM formulation \cite{Boyd2010}. Note that, given that the price impact of each aggregator affects everybody else, we cannot separate Eq.~(\ref{eq:OF_split}) in the variable $i$, \emph{i.e.} the equation is coupled and the sum's terms cannot be independently distributed among the aggregators. This type of problem is suited to be formulated as a \emph{global variable consensus problem} \cite{Boyd2010}, which works as follows. Consider a minimisation problem in the following form:
$$\min_{\mathbf{x}} \sum_{i=1}^n f_i(\mathbf{x})$$ where the goal is that each term in the sum can be handled independently. In the cases where the variable $\mathbf{x}$ is not separable in $i$, \emph{local} variables $\mathbf{x}^i$ and a \emph{global} variable $\mathbf{z}$ can be introduced, rewriting the problem as:
\begin{align*}
&\min_{\{\mathbf{x}^i\}} \sum_{i=1}^n f_i(\mathbf{x}^i)\\
\text{subject to:}\,\,\,& \mathbf{x}^i - \mathbf{z} = 0, \, \forall =1,\ldots,n
\end{align*}
As mentioned above, the problem constraints require all local variables to agree with each other and with the global variable. This way, global consensus on the solution is achieved. Also, note that $f_i$ uses only aggregator $i$'s individual constraints, which can be embedded into the function $f_i$ itself.

In a similar vein and focusing on our scenario, let $\mathbf{E}$ and $\mathbf{E}^{(i)}$ be the global and local variables respectively, each of which comprises a vector with dimension $24 n$ \emph{i.e.} $\mathbf{E}^{(i)} = \left( \mathbf{E}^{(i),1}, \ldots, \mathbf{E}^{(i),n} \right)$ and  $\mathbf{E}^{(i), j} = \left( E^{(i),j}_0, \ldots, E^{(i),j}_{23} \right)$. Following Eq.~(\ref{eq:OF_split}), the functions $f_i$ are given by:
\begin{equation*}
f_i\left( \mathbf{E}^{(i)}\right) = 
\begin{cases}
\sum_{t=0}^{23} \left[ E^{(i), i}_t \cdot \hat{\mathcal{P}}_t^{\text{convex}} \left( \sum_{j=1}^n E^{(i),j}_t \right) \right] &\hspace{-0.3cm} , \; \vspace{0.2cm}\text{\parbox{4cm}{if constraints (\ref{eq:PI1}), (\ref{eq:PI2}), (\ref{eq:PI3}), (\ref{eq:PI4}) are met by $\mathbf{E}^{(i),  i}$}}\\
\infty & \hspace{-0.3cm} , \;\text{otherwise}
\end{cases}
\end{equation*}

The resulting ADMM algorithm is then given by the following iterative equations:
\begin{subequations}
\begin{align}
& \mathbf{E}_{[k+1]}^{(i)} = \argmin_{\mathbf{E'}} \left( f_i(\mathbf{E'}) + {\boldsymbol{\xi}_{[k]}^{(i)\; T}} \cdot \left( \mathbf{E'} - \mathbf{E}_{[k]} \right) + \dfrac{\rho}{2} \norm{\mathbf{E'}-\mathbf{E}_{[k]}}_2^2 \right)  \label{eq:ADMM1}\\
& \mathbf{E}_{[k+1]} = \dfrac{1}{n} \sum_{i=1}^n \left( \mathbf{E}_{[k+1]}^{(i)} + \dfrac{1}{\rho} \boldsymbol{\xi}_{[k]}^{(i)} \right)  \label{eq:ADMM2}\\
& \boldsymbol{\xi}_{[k+1]}^{(i)} = \boldsymbol{\xi}_{[k]}^{(i)} + \rho\left( \mathbf{E}_{[k+1]}^{(i)} -  \mathbf{E}_{[k+1]} \right) \label{eq:ADMM3}
\end{align}
\end{subequations}
where the subscript $[k]$ denotes iteration number, and $\boldsymbol{\xi}$ and $\rho$ are the dual variable and the augmented Lagrangian parameter, respectively \cite{Boyd2010}. Intuitively, $\rho$ controls the trade-off between each aggregator solving its own local problem, and achieving global consensus (not necessarily to a minimum point). In more detail, if $\rho$ is set too high, the algorithm forces consensus \emph{too much}, resulting in very slow convergence. Conversely, if $\rho$ is set too small, each aggregator solves its local problem and consensus is not reached. Examples of this are presented in Section \ref{sec:cs_convergence}.

Given this, the iterative algorithm works as follows: first, each EV aggregator solves their local problem, Eq.~(\ref{eq:ADMM1}), and update their local copy of the energy schedule, $\mathbf{E^{(i)}}$. Then, an aggregation step, Eq.~(\ref{eq:ADMM2}), collects all the local solutions proposed by each aggregator and updates the global energy schedule, $\mathbf{E}$, reporting this vector back to all the aggregators. Lastly, each aggregator updates their local copy of the dual variable, $\boldsymbol{\xi}^{(i)}$, as per Eq.~(\ref{eq:ADMM3}) and proceeds to the new iteration.

This iterative process is stopped when the primal and dual residuals reach some user-specified tolerances, $\epsilon_{\text{pri}}$ and $\epsilon_{\text{dual}}$ \cite{Boyd2010,Munsing2017}. Specifically, the primal residual is denoted by $\mathbf{r}_{[k]} = \left( \mathbf{r}_{[k]}^1, \ldots, \mathbf{r}_{[k]}^n \right)$, where $\mathbf{r}_{[k]}^i = \mathbf{E}_{[k]}^{(i)} - \mathbf{E}_{[k]}$. Similarly, the dual residual is given by $\mathbf{s}_{[k]} = \mathbf{E}_{[k]} - \mathbf{E}_{[k-1]}$. The stopping criterion then takes the following form:
\begin{subequations}
\begin{align}
& \norm{\mathbf{r}_{[k]}}_2^2 \leq \epsilon_{\text{pri}} \label{eq:stop1}\\
& \norm{\mathbf{s}_{[k]}}_2^2 \leq \epsilon_{\text{dual}} \label{eq:stop2}
\end{align}
\end{subequations}
and the algorithm stops when both conditions have been met.

This concludes the exposition of the novel decentralised algorithm, which will be empirically tested in Section \ref{sec:cs_convergence}. We are now ready to study how the algorithm could be manipulated by a self-interested agent, and how this can be detected.

%%%%%%%%%%%%%%%%%%%%%%%%%%%%%%%%%%%%%%%%%%%%%%%%%%%%%%%%
%%%%%%%%%%%%%%%%%%%%%%%%%%%%%%%%%%%%%%%%%%%%%%%%%%%%%%%%
\section{Strategic Manipulation of the ADMM Algorithm}
\label{sec:attack}

The ADMM-based algorithm described in the previous section has nice convergence properties and (asymptotically) reaches the global optimum for suitable values of $\rho$ \cite{Boyd2010}. However, this requires every participating agent to run the algorithm faithfully. In our case, where agents are assumed to be self-interested, an aggregator could deviate from their assigned local algorithm and/or misreport their local solutions with the aim of improving their allocation. More specifically, in our scenario we assume a potential attacker aims to reduce its energy costs (\emph{i.e.} increase its utility). Therefore, in this section we focus on the strategic manipulation of our proposed ADMM algorithm, Eqs.~(\ref{eq:ADMM1}), (\ref{eq:ADMM2}), (\ref{eq:ADMM3}), and we will show how a misbehaving aggregator can significantly affect the algorithm's outcome. Note that we do not look at all possible manipulation vectors, as this is not feasible, but instead focus on several intuitive and specific types of manipulation that are beneficial for the attacker in our setting.

Formally, following the notation from Section \ref{sec:decOpt}, the electricity costs incurred by aggregator $i$ when a global allocation $\mathbf{E} = \left( \mathbf{E}^{1}, \ldots, \mathbf{E}^{n} \right)$ is reached are given by:
\begin{equation}
\text{cost}_{i} = \sum_{t=0}^{23} \left[ E^{i}_t \cdot \hat{\mathcal{P}}_t \left( \sum_{j=1}^n E^{j}_t \right) \right]
\label{eq:cost}
\end{equation}
In order to reduce these costs, the attacker aims to minimise the price impact on their desired hours, which in turn can be achieved by moving other aggregators' overlapping allocations to different hours. To this end, we consider different attack vectors, namely \emph{Shift}, \emph{Proportional}, \emph{Freeze}, \emph{FreezeShift},  \emph{FreezeProp} and \emph{Adversarial} attacks, which are described next. These capture different ways that an attacker can try to increase its own utility and present very different outcomes and efficacy, as detailed in Section \ref{sec:cs_attacks}. More specifically as described before, each attack will try to push the attacked aggregator's bids outside of the hours preferred by the attacker, in order to reduce price impact and hence the attacker's energy costs given by Eq.~(\ref{eq:cost}). Note that, throughout the rest of the paper, we assume that an attacker performs a given attack vector with a given intensity in every round. The study of more sophisticated attacks is outlined as future work (Section \ref{sec:conclusion}). For quick reference, a summary of the considered attack vectors can be found in Table \ref{table:attacks}. Finally, note that an empirical evaluation of all the proposed attack vectors can be found in Section \ref{sec:cs_attacks}.

\begin{table}[!t]
\centering
\begin{tabular}{|c|l|}
\hline
Attack name       & \multicolumn{1}{c|}{Short description}                                                                                                            \\ \hline
Shift(All)        & \begin{tabular}[c]{@{}l@{}}Shift the proposed allocation for one attacked aggregator\\ (or all aggregators) to more expensive hours\end{tabular}  \\ \hline
Proportional(All) & \begin{tabular}[c]{@{}l@{}}Scale down the proposed allocation for the attacked\\ aggregator (or all aggregators)\end{tabular}                     \\ \hline
Freeze            & \begin{tabular}[c]{@{}l@{}}Propose its individually optimal allocation for itself\\  (without considering the competitor aggregators)\end{tabular} \\ \hline
FreezeShift(All)  & Freeze + Shift(All)                                                                                                                               \\ \hline
FreezeProp(All)   & Freeze + Proportional(All)                                                                                                                        \\ \hline
Adversarial       & \begin{tabular}[c]{@{}l@{}}Attempt to incriminate a benign aggregator as deviator\\ by artifically favouring their allocations\end{tabular}     \\ \hline
\end{tabular}
\caption{Summary of the proposed attack vectors.}
\label{table:attacks}
\end{table}

%%%%%%%%%%%%%%%%%%%%%%%%%%%%%%%%%%%%%%%%%%%%%%%%%%%%%%%%
\subsection{Shift Attack}
\label{sec:shiftAttack}

\begin{figure}[t]
\centering
        \stackunder[5pt]{\includegraphics[width=.26\linewidth]{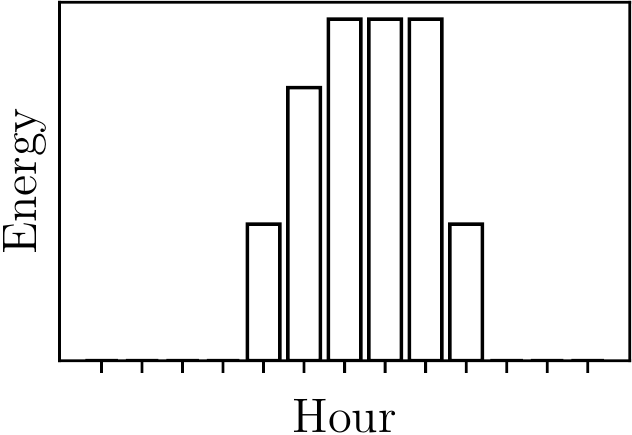}}{(a)}%
        \hspace{1cm}
        \stackunder[5pt]{\includegraphics[width=.26\linewidth]{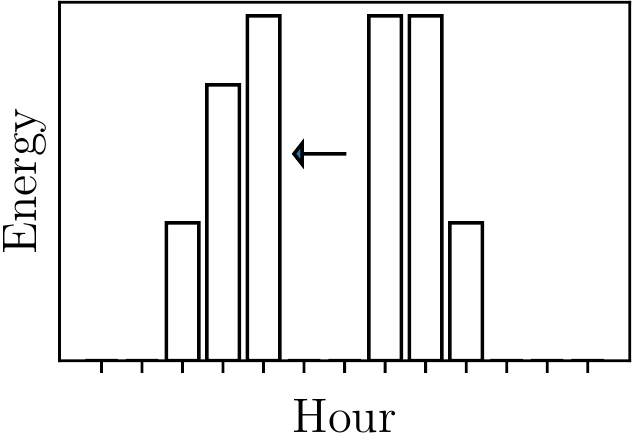}}{(b)}%
        \vspace{-.3cm}
        \caption{(a) Truthful allocation from aggregator $i$ to aggregator $j$ following Eq.~\ref{eq:ADMM2}, $\mathbf{E}_{[k+1]}^{(i),j}$. 
(b) Attacked $\hat{\mathbf{E}}_{[k+1]}^{(i),j}$ employing a shift attack with $\mu=2$ as given by Eq.~\ref{eq:shiftAttack}.}
        \label{fig:shiftAttack}
\end{figure}

In this type of attack, the deviating aggregator $i$ runs its local optimisation problem (\emph{i.e.} Eq.~\ref{eq:ADMM2}), but artificially modifies its local schedule allocation for another aggregator $j$, $\mathbf{E}^{(i), j}$, by \emph{shifting} it outside of aggregator $i$'s preferred hours. In this work and without loss of generality, we will focus on a particular case where the deviating aggregator splits the energy schedule of the attacked aggregator $j$ by its mid-hour, and then shifts the first half outwards by a number of hours (the \emph{strength} of the attack) $\mu=1,2,\ldots$, as depicted in Fig. \ref{fig:shiftAttack}. This is motivated by the fact that, normally, the cheapest prices lie somewhat in the middle hours of the day. Note that the analogous attack where both halves are shifted outwards was also considered and its empirical results were found to be very similar.

In more detail, let $t^*$ be the median hour with non-negative energy allocation for agent $j$, $\mathbf{E}_{[k+1]}^{(i),j}$. Then, given $\mathbf{E}_{[k+1]}^{(i)}$ from Eq.~\ref{eq:ADMM2}, the allocation of aggregator $j$ is modified as follows:
\begin{equation}
\hat{E}_{[k+1], \; t}^{(i),j} =
\begin{cases}
E_{[k+1], \; t+\mu}^{(i),j}\,\, , \,\, \text{if  } t\leq \floor*{t^*}-\mu \\
0 \,\, , \,\, \text{if  } t \in \left( \floor*{t^*}-\mu, \floor*{t^*} \right] \\
E_{[k+1], \; t}^{(i),j}\,\, , \,\, \text{if  } t > t^*
\end{cases}
\label{eq:shiftAttack}
\end{equation}
Note that, in the mathematical formulation presented in Eq.~(\ref{eq:shiftAttack}), the allocation can be pushed beyond the 24h day interval for large values of $\mu$, but this does not happen for the range of values employed in the empirical evaluation described in Section \ref{sec:experiments}.

Finally, this attack vector can be extended to target all the competing aggregators, rather than aggregator $j$ only, and will then be referred to as \emph{ShiftAll} attack.

\begin{figure}[t]
\centering
\includegraphics[width=.85\linewidth]{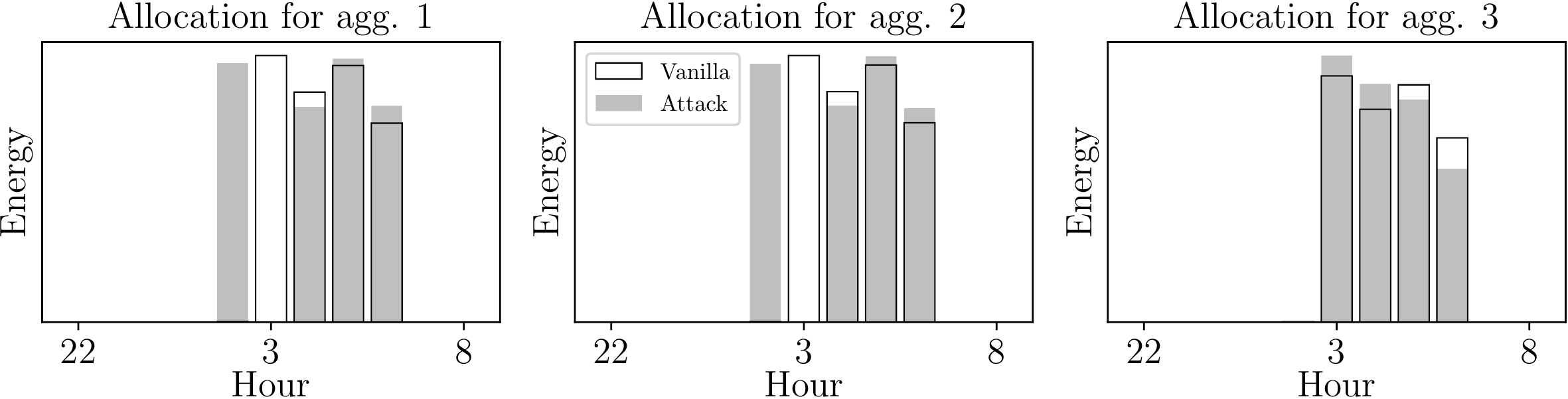}
\caption{Real example of the effects of \emph{ShiftAll} on the resulting energy allocations. The scenario consists of three aggregators of the same size with the third aggregator being the attacker and employing $\mu = 1$. Results for both vanilla and attacked scenarios are presented.}
\label{fig:exampleShift}
\end{figure}

For illustrative purposes, the effects of \emph{ShiftAll} in a scenario with three aggregators are shown in Fig. \ref{fig:exampleShift}. The third aggregator deviates from the vanilla algorithm and manages to shift the allocation of the first two aggregators out of one of its preferred hours (3 am). As a result, the attacker is able to obtain more energy at 3 am and reduce its allocation during the more expensive 
hours 5 and 6 am.

%%%%%%%%%%%%%%%%%%%%%%%%%%%%%%%%%%%%%%%%%%%%%%%%%%%%%%%%
\subsection{Proportional Attack}
\label{sec:proportionalAttack}

In this type of manipulation, the deviating aggregator $i$ runs its local optimisation problem (\emph{i.e.} Eq.~\ref{eq:ADMM2}), but scales down its allocation for another aggregator $j$, $\mathbf{E}^{(i), j}$, by a factor $\lambda \in [0, 1]$, which indicates the \emph{strength} of the attack. Formally, $\mathbf{E}_{[k+1]}^{(i)}$ is obtained from Eq.~\ref{eq:ADMM2}, and then modified as:
\begin{equation}
\hat{\mathbf{E}}_{[k+1]}^{(i),j} = \mathbf{E}_{[k+1]}^{(i),j} \cdot \left( 1-\lambda \right)
\label{eq:proportionalAttack}
\end{equation}
The effect of this attack is to flatten the the allocation of aggregator $j$, resulting in less overlap with aggregator $i$'s desired schedule, in a similar way to the \emph{Shift} attack. Note that the attacked aggregator still enforces its own constraints, Eqs. (\ref{eq:PI1})-(\ref{eq:PI4}), so that the amount of energy it receives is enough to satisfy its requirements.

\begin{figure}[!t]
\centering
\includegraphics[width=.85\linewidth]{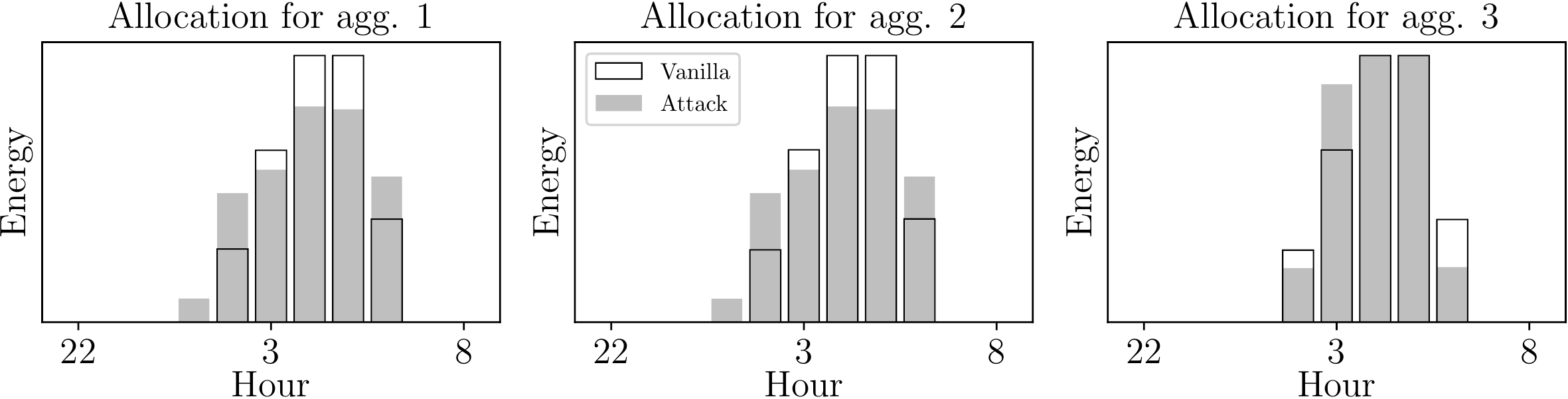}
\caption{Real example of the effects of \emph{ProportionalAll} on the resulting energy allocations. The scenario consists of three aggregators of the same size with the third aggregator being the attacker and employing $\lambda = 0.66$. Results for both vanilla and attacked scenarios are presented.}
\label{fig:exampleProp}
\end{figure}

Analogously to the previous attack vector, \emph{Proportional} can be targetted to all the competing aggregators. In such a case it will be denoted by \emph{ProportionalAll}. As an example, the effects of \emph{ProportionalAll} in a scenario with three aggregators are shown in Fig. \ref{fig:exampleProp}. The third aggregator deviates from the vanilla algorithm and manages to flatten the allocation of the first two aggregators out of one of its preferred hours (3, 4, 5 am). As a result, the attacker is able to obtain more energy at the cheap hours and reduce its consumption during more expensive ones.

%%%%%%%%%%%%%%%%%%%%%%%%%%%%%%%%%%%%%%%%%%%%%%%%%%%%%%%%
\subsection{Freeze Attack}
\label{sec:freezeAttack}

In this case, attacker $i$ \emph{freezes} its own allocation to the individually optimal one, \emph{i.e.} the allocation that can be obtained by solving Eqs.~(\ref{eq:PI}), (\ref{eq:PI1}), (\ref{eq:PI2}) and (\ref{eq:PI3}) without taking into account the other aggregators. Formally, being $\mathbf{E}^*_i$ the optimal individual allocation for aggregator $i$:
\begin{equation}
\hat{\mathbf{E}}_{[k+1]}^{(i),i} = \mathbf{E}^*_i
\label{eq:freezeAttack}
\end{equation}
This way, the attacker hopes to get its individually optimal allocation and get the other competing aggregators to arrange their allocations around. Importantly, this attack vector can be combined with the others presented so far. Specifically, we call the attack vectors combining Eqs.~(\ref{eq:shiftAttack}) and (\ref{eq:freezeAttack}) \emph{FreezeShift} and \emph{FreezeShiftAll}. Similarly, the attacks combining Eqs.~(\ref{eq:proportionalAttack}) and (\ref{eq:freezeAttack}) will be called \emph{FreezeProp} and \emph{FreezePropAll}. Intuitively, these combinations should yield more benefit to the attacker since its own allocation will be less affected by the smoothing effect of the competing aggregators.

%%%%%%%%%%%%%%%%%%%%%%%%%%%%%%%%%%%%%%%%%%%%%%%%%%%%%%%%
\subsection{Adversarial Attack}
\label{sec:adversarialAttack}

This last type of attack we consider is different from the previously described ones, as the deviating aggregator $i$ does not seek to directly manipulate another aggregator's allocation. Instead, it will try to incriminate a benign aggregator $j$ to make it appear as a deviator, hoping it will be a false positive of the manipulation detection algorithm (discussed in Section \ref{sec:detectManipulation}) and penalised accordingly. Depending on the imposed penalty, this could consist on banning aggregator $j$'s participation on the current trading day, thus benefiting aggregator $i$ as competition is reduced. Otherwise, this can be seen as a purely malignant adversarial attack. Note that this sort of attacks have attracted a lot of recent interest in the field of machine learning \cite{Huang2011,Kurakin2016} and could present a serious drawback for incentivising agents to participate in cooperative schemes such as the one proposed in this work.

Formally, this attack can be performed by proposing a schedule for aggregator $j$, $\mathbf{E}_{[k+1], \; t}^{(i),j}$, artificially close to the allocation of aggregator $j$ to itself in the previous round, $\mathbf{E}_{[k]}^{(j),j}$. Hence aggregator $j$ appears to deviate from the algorithm as it breaks the balance in aggregators $i$-$j$ interaction. This will become clear in Section \ref{sec:quantManipulation} where we describe how to quantify manipulation. This attack can be parametrised by a parameter $\lambda \in [0,1]$ which determines a linear combination between the schedule proposed by aggregator $j$ to itself, and the schedule allocated by aggregator $i$ to $j$ as a result of Eq.~(\ref{eq:ADMM1}). Formally, given $\mathbf{E}_{[k+1]}^{(i),j}$ from Eq.~\ref{eq:ADMM2}, modify the allocation to aggregator $j$ as follows:
$$\hat{\mathbf{E}}_{[k+1]}^{(i),j} = \mathbf{E}_{[k+1]}^{(i),j} \cdot \left( 1-\lambda \right) + \mathbf{E}_{[k]}^{(j),j} \cdot \lambda$$
In more detail, an attack with parameter $\lambda = 1$ proposes an allocation to aggregator $j$ equal to what $j$ proposed for itself in the previous round. This is likely to be beneficial for aggregator $j$'s schedule, as it will contribute towards maintaining the more favourable schedules characteristic of early rounds before convergence. However, as will be detailed in Section \ref{sec:quantManipulation}, this will make benign aggregator $j$ seem a deviator, with the subsequent penalty. Conversely, as $\lambda$ tends to zero, we recover the vanilla ADMM algorithm.

%%%%%%%%%%%%%%%%%%%%%%%%%%%%%%%%%%%%%%%%%%%%%%%%%%%%%%%%
%%%%%%%%%%%%%%%%%%%%%%%%%%%%%%%%%%%%%%%%%%%%%%%%%%%%%%%%

\section{Detecting Manipulation}
\label{sec:detectManipulation}

In this section, we detail a mathematical framework for quantifying the influence of a given ADMM participant, \emph{i.e.} an EV aggregator, onto the rest of participants. The aim is to be able to detect outliers that are symptom of strategic manipulation in the system. Although this framework is general, and can be applied to any ADMM (or variant) scenario, we focus on our particular case for ease of exposition.

%%%%%%%%%%%%%%%%%%%%%%%%%%%%%%%%%%%%%%%%%%%%%%%%%%%%%%%%

\subsection{Quantifying Manipulation}
\label{sec:quantManipulation}

The basic idea is that any group of aggregators with overlapping energy requirements should influence each other's schedules with \emph{similar} intensity. If a particular aggregator $i$ is self-interested and wants to improve its allocation by deviating from the ADMM algorithm, it will exert a heavier influence onto its competitors' allocations. Conversely, as happens in the adversarial attack detailed in Section \ref{sec:adversarialAttack}, an aggregator that tries to wrongly flag another benign aggregator as deviator would exert too little influence.

A key point is that each aggregator $i$ produces a (local) proposed schedule for all the $n$ participating aggregators. Formally, following the notation from Section \ref{sec:decOpt}:
$$\mathbf{E}_{[k+1]}^{(i)} = \left( \mathbf{E}_{[k+1]}^{(i),1}, \ldots, \mathbf{E}_{[k+1]}^{(i),n} \right)$$
Hence, this local solution proposed by aggregator $i$ at iteration $k$ contains its own schedule, $\mathbf{E}_{[k+1]}^{(i), i}$, and all the schedules for all the other participants, $\mathbf{E}_{[k+1]}^{(i),j}$ for $j\neq i$. We assume that each aggregator, benign or deviator, is truthful about their own allocations in their proposed local solutions. The reason for this is that every aggregator wants the best energy schedule given their requirements and hence would report the optimal schedule arising from their minimisation problem. Moreover, note that, given that an aggregator is unlikely to frequently change size and needs electricity everyday, past behaviour can also be used to roughly infer some of the aggregator's characteristics. The study of more general manipulation settings is out of the scope of this paper and is outlined as future work in Section \ref{sec:conclusion}. Also, without loss of generality, we assume that the deviating behaviour starts from the second ADMM round, when every aggregator has seen the proposals from each aggregator. This allows us to focus on the first two iterations ($k=0,1$) for ease of exposition.

\begin{figure}[!t]
\centering
\includegraphics[width=.9\linewidth]{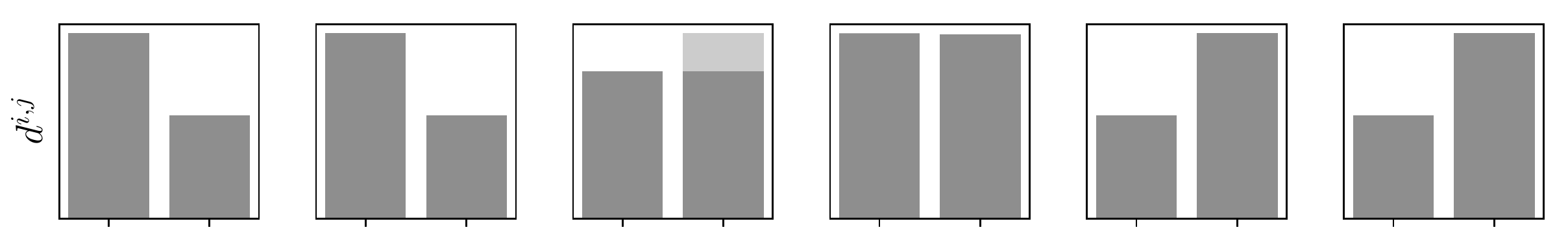}
\includegraphics[width=.9\linewidth]{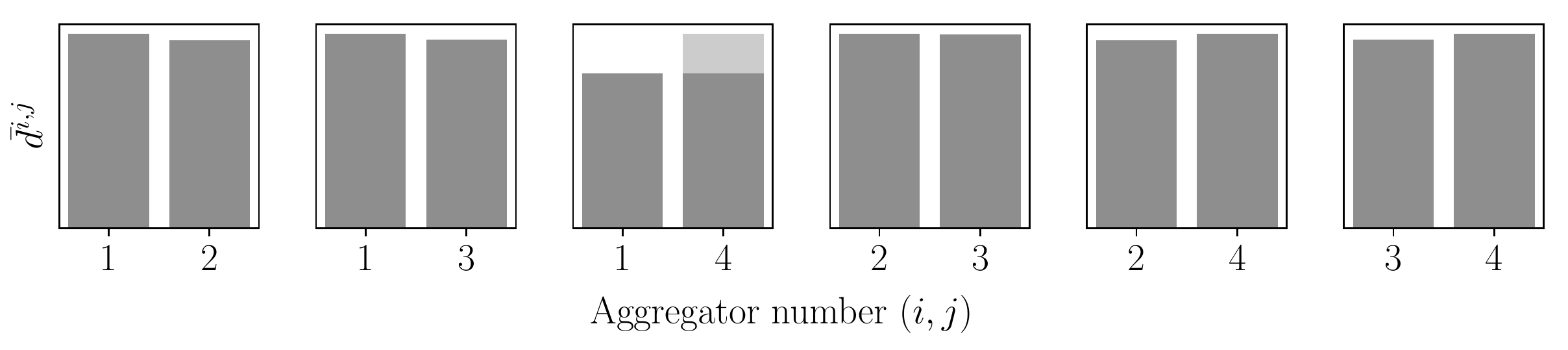}
\vspace{-0.5cm}
\caption{\emph{Difference matrix} (top) and \emph{normalised difference matrix} (bottom), $d$ and $\bar{d}$ respectively, for a scenario with two aggregators of size $50\; 000$ EVs (1 and 4) and two aggregators of size $150\;000$ EVs (2 and 3). One of the small aggregators (Aggregator 4) performs a \emph{Shift} attack with $\mu=1$ against the other small aggregator (Aggregator 1), displayed as light grey.}
\label{fig:normalisation}
\end{figure}

Formally, let $d$ be a square matrix of dimension $n$, the \emph{difference matrix}, storing how much each aggregator affects its competitors' self-proposed allocations. In more detail, every $i,j$ entry quantifies how much aggregator $i$ modifies the self-assigned schedule of agent $j$, and is given by:
$$d^{i,j} = \lVert E^{(i), j}_{[1]} - E^{(j), j}_{[0]} \rVert$$
As mentioned above, we expect benign aggregators to affect each other's schedules in a similar way, but aggregator size significantly affects this. More precisely, there are natural magnitude deviations in $d^{i,j}$ and $d^{j,i}$ when the sizes of the benign aggregators $i$ and $j$ differ. An example of this effect is shown in the top row of Fig. \ref{fig:normalisation} in dark grey.

These natural differences are an obstacle for detection algorithms. In order to overcome this issue, we normalise the matrix $d$ employing the total amount of energy allocated by each aggregator to itself, as a proxy to potentially unknown aggregator size. Note that other proxies can be used instead, such as actual size and total amount of electricity purchased in previous trading days. Formally, we can write:
$$\text{size}_i = \sum_{t=0}^{23} E_{[0]}^{(i), i}$$
and the proportion of the size of aggregator $i$ among the whole group of aggregators is given by:
$$p_i = \dfrac{\text{size}_i}{\sum_j{\text{size}_j}}$$
Then, the \emph{normalised difference matrix}, $\bar{d}$, is given by:
\begin{equation}
\bar{d}^{i,j} = \lVert E^{(i), j}_{[1]} - E^{(j), j}_{[0]} \rVert \cdot \dfrac{\sqrt{p_i}}{\text{size}_i + \text{size}_j}
\label{eq:dbar}
\end{equation}
This scaling function was chosen as it empirically \emph{flattens} the entries of the matrix $\bar{d}$ corresponding to benign aggregators, eliminating most of the dependence on aggregator size. In more detail, extensive simulations were performed, studying a variety of scenarios with different number of aggregators of different sizes, and different attack vectors and attack strengths. The selected  normalisation approach provides the best results. An example of the normalisation effect is pictured in the bottom row of Fig. \ref{fig:normalisation}. In this plot, the effect of the manipulating aggregator is shown in light grey, whereas the rest of the dark grey bars correspond to benign behaviour. In the top plots, corresponding to the \emph{difference matrix} $d$, we can see large differences between different entries, arising from the large size differences between the aggregators. Importantly, these natural differences are larger that the effect of the manipulating aggregator. In contrast, the normalised bottom plots manage to nearly flatten all the natural differences, and the effect of the deviating aggregator clearly stands out.

Lastly, for the rest of the paper, we assume that $n-1$ aggregators are benign and only one of them can potentially be a deviator. This is motivated by the fact that, with a perfect detection algorithm, there exists a Nash equilibrium in which no-one wants to deviate. Note that the proposed detection algorithm, which we are now ready to introduce, could be extended to deal with the more general case of having any number of deviators.

%%%%%%%%%%%%%%%%%%%%%%%%%%%%%%%%%%%%%%%%%%%%%%%%%%%%%%%%%%%%%%%%%%%%%%%%
\subsection{Detecting Manipulation}
\label{sec:detManipulation}

\begin{algorithm}[t]
\SetKwInOut{Input}{Input}\SetKwInOut{Output}{Output}

 \Input{$\bar{d}, \alpha$}
 \Output{list with the detected manipulating aggregator, if any}
 
 \BlankLine \BlankLine
 \tcc{off-diagonal}
 consider the off-diagonal elements: \texttt{offDiag}\;
 compute the median: $\mu_{1/2} = median( \texttt{offDiag})$\;
 compute distances from each element in \texttt{offDiag} to $\mu_{1/2}$\;
 find max distance $\longrightarrow$ \texttt{maxOffDiag}\;
 
 \BlankLine \BlankLine
 \tcc{on-diagonal}
 consider the on-diagonal elements: \texttt{onDiag}\;
 compute the median: $\mu_{1/2} = median( \texttt{onDiag})$\;
 compute distances from each element in \texttt{onDiag} to $\mu_{1/2}$\;
 find max distance $\longrightarrow$ \texttt{maxOnDiag}\;
 
 \BlankLine \BlankLine
 \tcc{threshold-based detection}
 max(\texttt{maxOffDiag}, \texttt{maxOnDiag}) $\longrightarrow$ \texttt{max}\;
 aggregator index: index(\texttt{max}) $\longrightarrow i$\;
   \eIf{\upshape{\texttt{max}} $> \alpha$}{
   deviator $\longrightarrow [i]$\;
   }{
   deviator$\longrightarrow []$\;
  }
  \BlankLine \BlankLine
 \Return{\upshape{deviator}}
 \caption{Threshold-based strategic manipulation detection algorithm for a scenario with at most one deviator.}
 \label{alg:detection}
\end{algorithm}

The overall idea is to be able to detect deviating aggregators in order to penalise them and discourage manipulation. As it is usually the case in complex stochastic environments, the aim here is to reduce false positives and false negatives, while keeping true positives and true negatives as high as possible. Specifically, in this work we consider a \emph{positive} to be an aggregator detected as deviator, and a \emph{negative} an aggregator classified as benign.

As explained in previous sections, the idea is that manipulating behaviour will stand out, as it exerts a larger or smaller influence in other aggregators allocations, compared to the scenario's average. Formally, one can use the normalised difference matrix $\bar{d}$ defined in the previous section in order to quantify this mathematically: manipulating behaviour from aggregator $i$ towards aggregator $j$ is translated into a too large or too small entry $\bar{d}^{i,j}$. As a first step towards manipulation detection, we propose applying a threshold-based algorithm, as described in Algorithm \ref{alg:detection}. In more detail, the algorithm looks at the difference matrix $\bar{d}$, computes the medians of the matrix entries, and then finds the entry that deviates the most from the median. This is done separately for off- and on-diagonal elements (as there are intrinsic magnitude differences between $\bar{d}^{i,i}$ and  $\bar{d}^{i,j}$ even when all aggregators are benign) and only the highest deviation of the two is taken as final candidate. Lastly, this candidate is classified as deviator if its deviation from the median is greater than the user-defined threshold $\alpha$.

The choice of threshold $\alpha$ is critical and we empirically study the performance of different thresholds in Section \ref{sec:cs_thresholds}. Also, although the presented algorithm is designed to work in scenarios with at most one manipulating agent, by selecting the aggregator that deviates the most, it can be easily adapted to a general scenario. The most straightforward way would be to simply classify as deviator any aggregator $i$ with $|\mu_{1/2} - \bar{d}^{i,j}| > \alpha$ for some $j$. This extended algorithm is conceptually the same as Algorithm \ref{alg:detection} and will be studied in future work.

Finally, we would like to note that the proposed algorithm could be used in conjunction with other detection methods in order to provide better results. We are now ready to present an empirical evaluation in order to test the performance of the decentralised algorithm proposed in Section \ref{sec:decOpt}, the different attack vectors proposed in Section \ref{sec:attack} and the manipulation detection algorithm presented in this section.

%%%%%%%%%%%%%%%%%%%%%%%%%%%%%%%%%%%%%%%%%%%%%%%%%%%%%%%%%%%%%%%%%%%%%%%%
%%%%%%%%%%%%%%%%%%%%%%%%%%%%%%%%%%%%%%%%%%%%%%%%%%%%%%%%%%%%%%%%%%%%%%%%
\section{Empirical Evaluation}
\label{sec:experiments}

In this section we present an analysis of the performance of the decentralised coordination algorithm (Section \ref{sec:decOpt}), the different attack vectors (Section \ref{sec:attack}) and the manipulation detection framework (Section \ref{sec:detectManipulation}). This empirical evaluation uses real market and vehicle usage data from Spain. We will start by detailing the scenario used in the simulations, and then describe the empirical results.

%%%%%%%%%%%%%%%%%%%%%%%%%%%%%%%%%%%%%%%%%%%%%%%%%%%%%%%%%%%%%%%%%%%%%%%%
\subsection{Experimental Setup}
\label{sec:experimentalSetup}

The experiment setup described in this section closely follows the case studies presented in \cite{Perez-Diaz2018a,Perez-Diaz2018b,Perez-Diaz2018d}. We consider a night-time residential scenario in which EVs arrive in the evening and need to be charged by the next morning. The EVs are assumed to be medium-sized with $24 \text{kWh}$ battery capacity and maximum charging speed $P_{\text{max}}= 3.7 \text{kW}$. Moreover, charging efficiency is set to $90\%$.

Real market data from the Spanish day-ahead market OMIE\footnote{\url{http://www.omie.es/en/inicio}} is used in the simulations, as described in \cite{Perez-Diaz2018a}. Specifically, for this paper we focus on trading data from November 2016.
Similarly, real driver data from a Spanish survey is used to determine probabilistic EV driving patterns, as detailed in \cite{Perez-Diaz2018a}. In more detail, we employ the distribution of times for the first and last trip from and to home, as shown in Table \ref{table:times}.

\begin{table}[!t]
\centering
\begin{tabular}{|p{.28cm}|p{1.7cm}|p{.62cm}|p{.62cm}|p{.62cm}|p{.62cm}|p{.62cm}|}
\hline
\multirow{2}{*}{$t_0$} & Time & 19h  & 20h  & 21h  & 22h  & 23h  \\ \cline{2-7}
                                      & Probability & 0.16 & 0.25 & 0.32 & 0.12 & 0.15 \\ \hline
\end{tabular}
\begin{tabular}{|p{.28cm}|p{1.7cm}|p{.62cm}|p{.62cm}|p{.62cm}|p{.62cm}|p{.62cm}|}
\hline
\multirow{2}{*}{$t_d$} &  Time & 6h    & 7h    & 8h    & 9h   & 10h \\ \cline{2-7}
                                             & Probability & 0.04  & 0.02  & 0.34  & 0.5  & 0.1\\   \hline
\end{tabular}
\caption{Possible arrival ($t_0$) and departure ($t_d$) times rounded to the nearest hour, with their respective probabilities.}
\label{table:times}
\end{table}

Regarding energy requirements, the desired state of charge of an EV at arrival and departure times are drawn from uniform distributions as follows: SoC$_0 \in \left[ \text{SoC}_\text{total}/4, \text{SoC}_\text{total}/2 \right]$ and 
SoC$_f \in \left[ 2\cdot \text{SoC}_\text{total}/3, \text{SoC}_\text{total} \right]$. Consequently, the EV charging requirements range between a large percentage of the battery (up to 75\%), to a small percentage (down to 16\%), accounting for long and short trips home.

%%%%%%%%%%%%%%%%%%%%%%%%%%%%%%%%%%%%%%%%%%%%%%%%%%%%%%%%%%%%%%%%%%%%%%%%
\subsection{Algorithm Convergence Results}
\label{sec:cs_convergence}

We start our experimental analysis by considering the convergence to the optimal solution of the proposed decentralised algorithm without any manipulation. A key determinant of convergence is the augmented Lagrangian parameter $\rho$ (see Eqs.~(\ref{eq:ADMM1}), (\ref{eq:ADMM2}),  (\ref{eq:ADMM3})). Intuitively, it controls the \emph{weight} that the similarity of local and global solutions has in the local minimisation algorithms (see Eq.~\ref{eq:ADMM1}). If it is set too large or too small, the algorithm will not converge. For every problem there exists a range of values providing convergence but, as already mentioned, it can be very slow in some cases. Also, the number of participating aggregators affects the convergence of the algorithm: the higher the number of participants, the more fragmented the optimisation problem is, so more iterations may be required. Thus, a suitable value for $\rho$ needs to be found in order to make the algorithm converge fast, a key point for its practical applicability.

\begin{figure}[!ht]
    \centering
    \includegraphics[width=.99\textwidth]{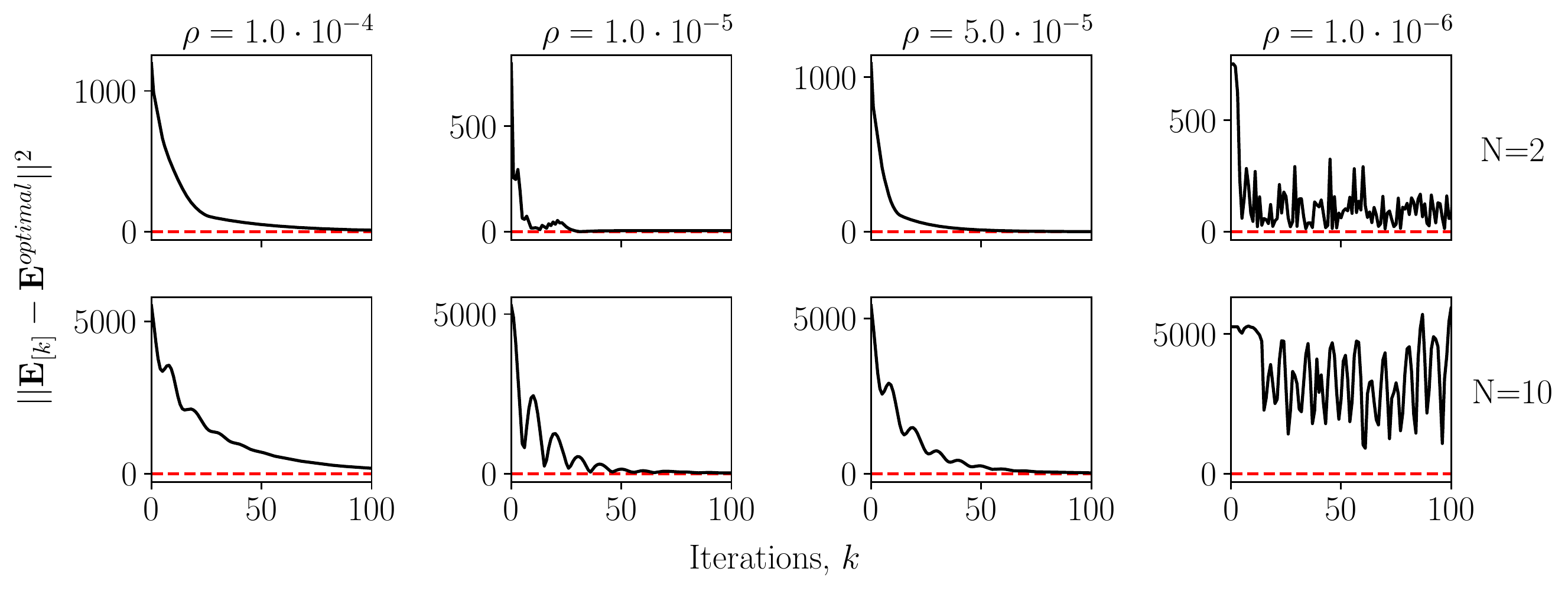} % first figure itself
    \vspace{-0.6cm}
    \caption{Convergence of the ADMM decentralised algorithm to the optimal centralised solution, for different values of $\rho$. (Top) Simulations with two  aggregators, each with $150\, 000$ EVs. (Bottom) Simulations with ten aggregators, each with $150\, 000$ EVs. Market data from 1/11/2016.}
    \label{fig:ADMMconvergence}
\end{figure}

Fig. \ref{fig:ADMMconvergence} shows the convergence for different values of $\rho$, and for two scenarios, with two and ten EV aggregators respectively. Each plot shows how far the decentralised solution is from the optimal solution. We can see similar convergence behaviour for the two scenarios, although the case with two EV aggregators is faster and more uniform. These results show evidence of good computational scaling with the number of EV aggregators, something key for tackling larger problem sizes.  Moreover, for both scenarios, convergence starts slow for a value of $\rho \lesssim 10^{-4}$, becoming fastest for a value $\rho \sim 10^{-5}$, and diverging for larger values. This suggests that a value about $\rho = 10^{-5}$ presents the best convergence for these scenarios, although this may vary for larger problem sizes. Also, we would like to note that these results are consistent across different trading days. Lastly, there exist recent extensions of the ADMM algorithm which include an adaptive parameter $\rho$ and can provide faster convergence and rule out the need for parameter tweaking \cite{Xu2017}.

%%%%%%%%%%%%%%%%%%%%%%%%%%%%%%%%%%%%%%%%%%%%%%%%%%%%%%%%%%%%%%%%%%%%%%%%
\subsection{Attack Vectors: Utility and Convergence}
\label{sec:cs_attacks}

\begin{figure}[!htb]
    \centering
    \begin{minipage}{0.495\textwidth}
        \centering
        \includegraphics[width=0.95\textwidth]{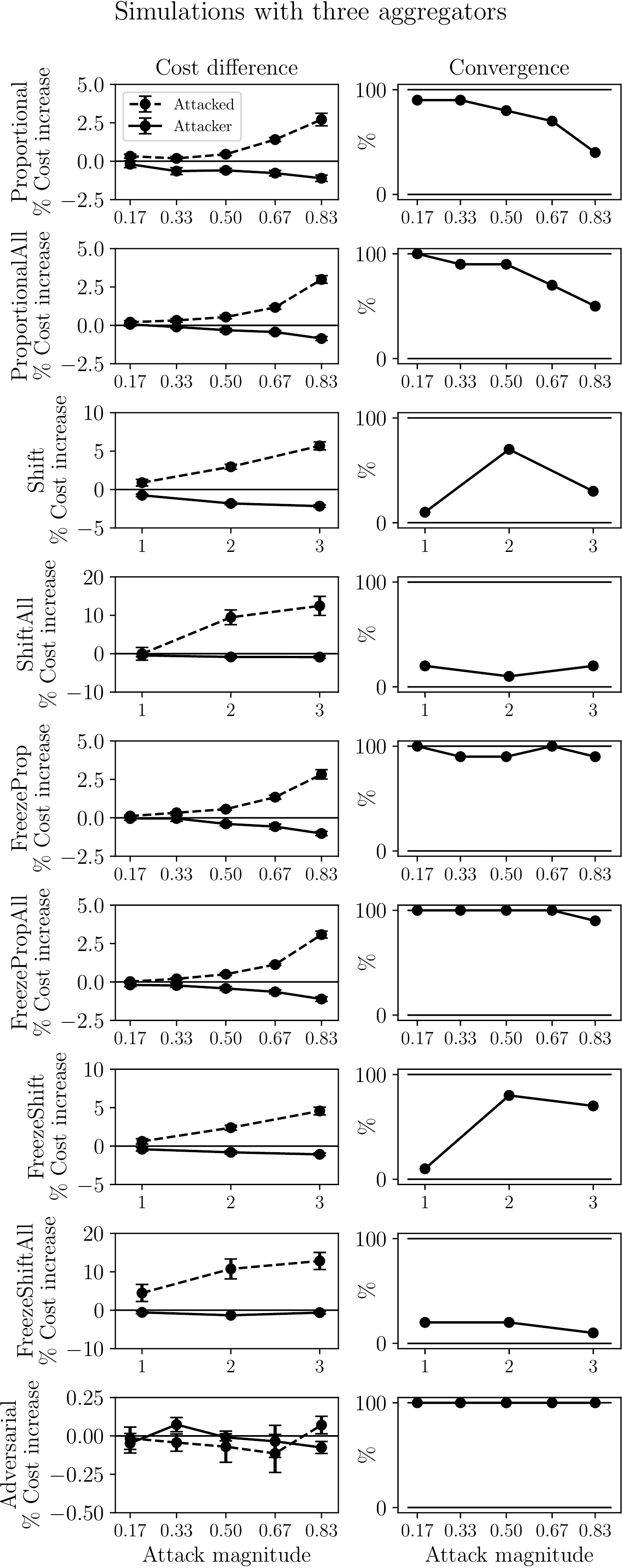}
    \end{minipage}%
    \begin{minipage}{0.495\textwidth}
        \centering
        \includegraphics[width=0.95\textwidth]{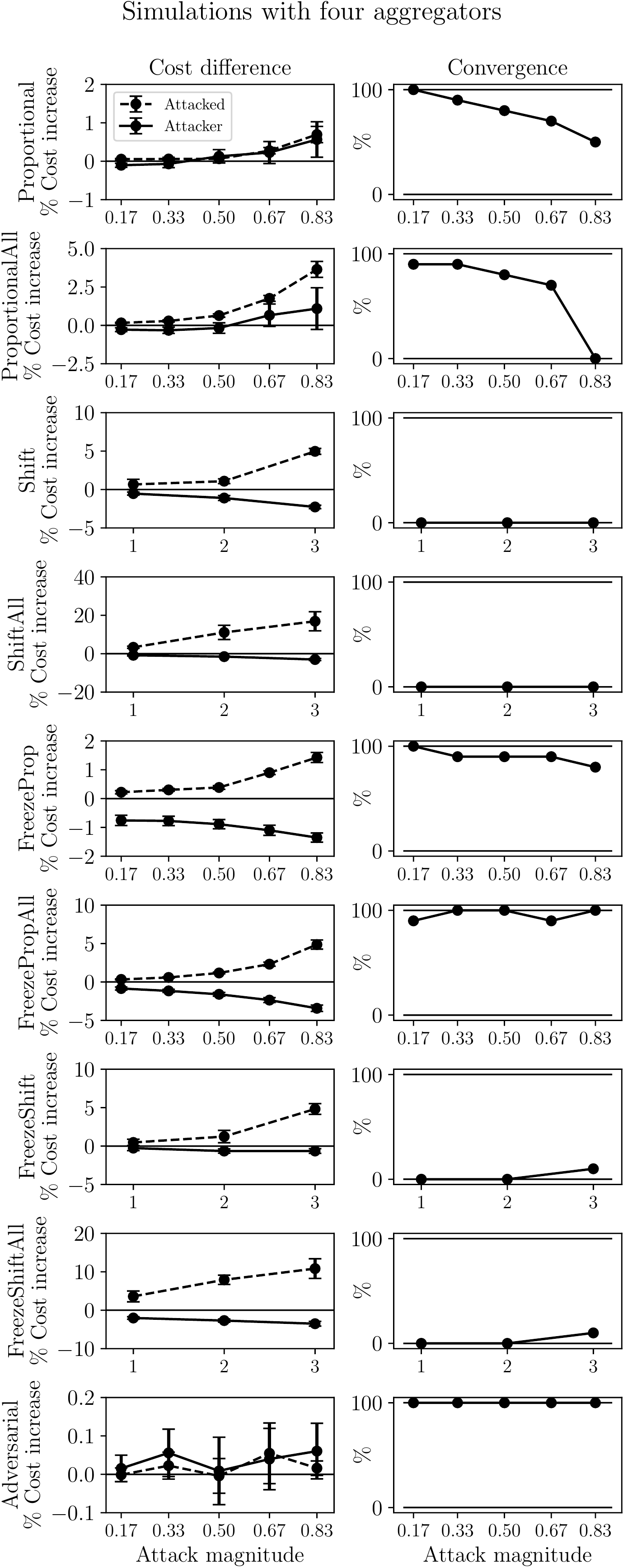}
    \end{minipage}
    \caption{Cost and convergence analysis for each of the attack vectors described in Section \ref{sec:attack}. Scenarios with three aggregators (LHS) and four aggregators (RHS) and averaged over the ten first days of November 2016. All aggregators have capacity for $150\;000$ EVs.}
    \label{fig:attacks}
\end{figure}

Similarly to the analysis presented in the previous section, where we studied the convergence of the proposed ADMM algorithm, we will now turn our attention to analysing the behaviour of the algorithm under the effect of the manipulative attacks presented in Section \ref{sec:attack}. As described in Section \ref{sec:lit}, existing works in the literature focus on attacks to the ADMM algorithm that seek to destabilise it and prevent its convergence. While this can beneficial for a malignant external attacker whose objective is to prevent the algorithm's operation, it is not necessarily so in our case with self-interested participants. In more detail, a given EV aggregator may want to completely prevent coordinated bidding by interrupting convergence of the proposed coordination algorithm, but most likely it will try to improve its own allocation by manipulating the algorithm in a subtle way that goes unnoticed. Consequently, we identify two key quantities to analyse in order to assess the efficacy of a given attack vector: the effectiveness of the considered attack, in terms of utility increase (\emph{i.e.} energy cost reduction) for the attacker, and the convergence of the algorithm under attack. In more detail, by convergence we refer to whether the primal and dual residuals decrease in successive iterations and the stopping criteria given by Eqs. \ref{eq:stop1} and \ref{eq:stop2} are eventually met. Moreover, the cost reductions are computed after 50 iterations irrespective of whether the attack converges or not. Note that this number of iterations is sufficient for convergence if the attack is successful.

In order to evaluate these two quantities, we run simulations for each of the proposed attack vectors over the ten first days of November 2016 using $\rho=10^{-5}$, and present the results in Fig. \ref{fig:attacks}. Here, we can see that both \emph{Proportional} and \emph{ProportionalAll} manage to achieve reduced costs for the attacker with good convergence rates in the scenarios with three aggregators. However, for settings with four aggregators, both these attacks destabilise the algorithm which outputs non-optimal allocations with increased costs for all the aggregators. Interestingly, both \emph{FreezeProp} and \emph{FreezePropAll} present very good results, consistently providing reduced costs for the attacker and close to $100\%$ convergence rates for all attack strengths. On the other hand, \emph{Shift}, \emph{ShiftAll}, \emph{FreezeShift} and \emph{FreezeShiftAll} present very large cost reductions but fail to converge in the vast majority of cases. We would like to note that even a $1\%$ cost reduction in the scenarios considered in this work represents savings in the order of hundreds of thousands of Euros per year. Finally, in a slightly different vein, \emph{Adversarial} presents $100\%$ convergence rates for all attack strengths and very small cost alterations. Recall that the aim of this attack is not to directly increase the attacker's utility, but to incriminate a benign aggregator as deviator (see Section \ref{sec:adversarialAttack}). As a consequence, the efficacy of \emph{Adversarial} is more appropriately measured by looking at the number of false positives (benign aggregators incorrectly classified as deviators) that the attack is able to generate. Results from this analysis are detailed in Section \ref{sec:adversarialAnalysis}, as they require parts of the explanation about the performance of the proposed detection algorithm, which is detailed in the next section.

Summarising, we have studied the effects of each of the attack vectors on the convergence and outcome of the proposed ADMM algorithm. While some attacks (\emph{Shift} and variants) prevent the algorithm from converging thus limiting their practical applicability, others, particularly \emph{FreezeProp} and \emph{FreezePropAll} are successful in decreasing the attacker's electricity costs while barely affecting the algorithms convergence. We would like to note that these interesting results all arise from myopic attack vectors that are simply repeated iteration after iteration. Given that an attacker is potentially able to monitor convergence and other metrics during successive iterations of the algorithm, it is likely that greater cost savings and convergence rates could be obtained by more sophisticated algorithms.

%%%%%%%%%%%%%%%%%%%%%%%%%%%%%%%%%%%%%%%%%%%%%%%%%%%%%%%%%%%%%%%%%%%%%%%%
\subsection{Threshold-Based Detection Results}
\label{sec:cs_thresholds}

\begin{figure}[htb]
\centering
\includegraphics[width=.9\linewidth]{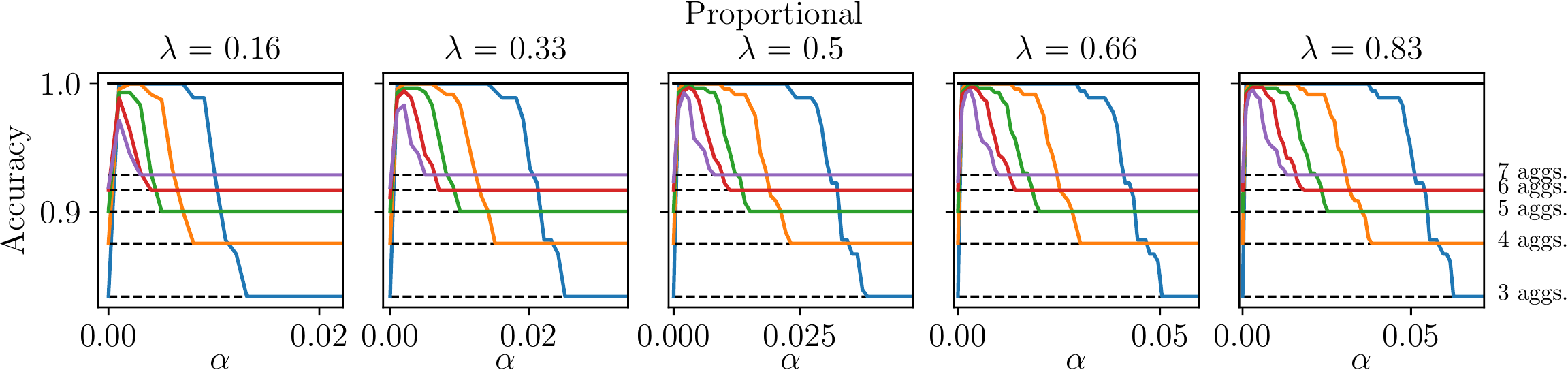}
\includegraphics[width=.58\linewidth]{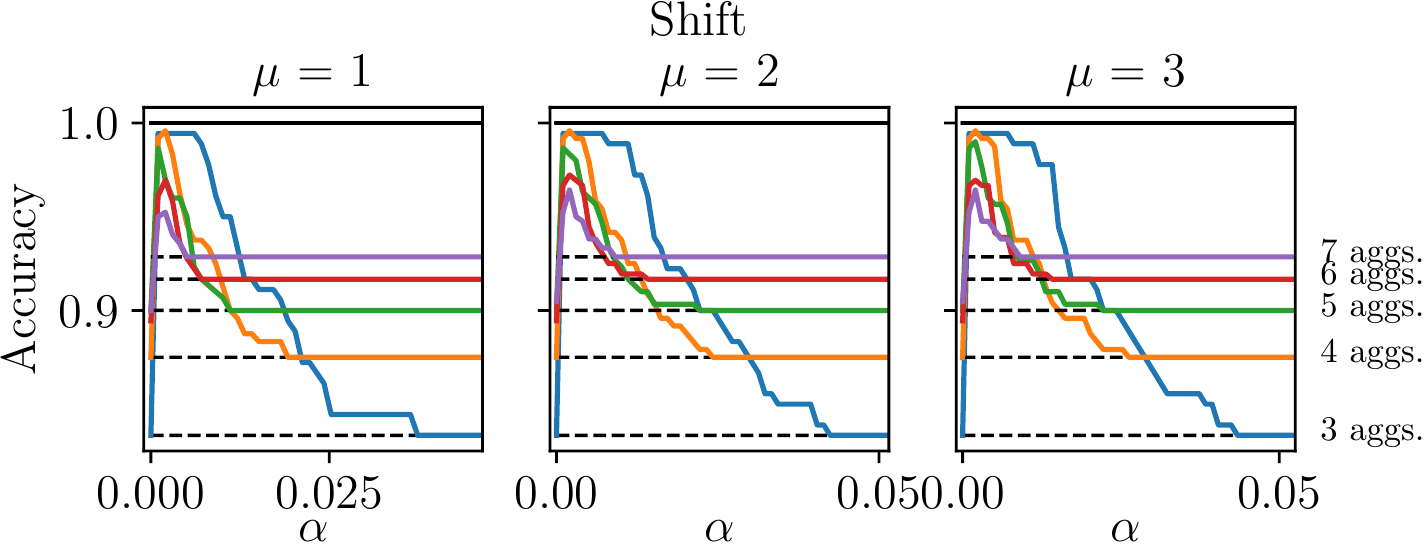}
\includegraphics[width=.9\linewidth]{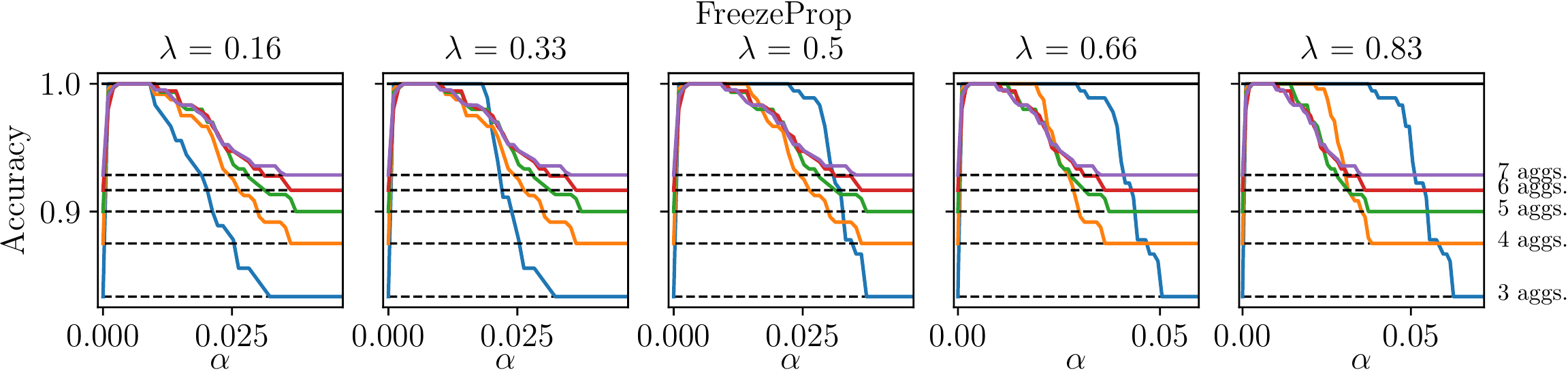}
\includegraphics[width=.58\linewidth]{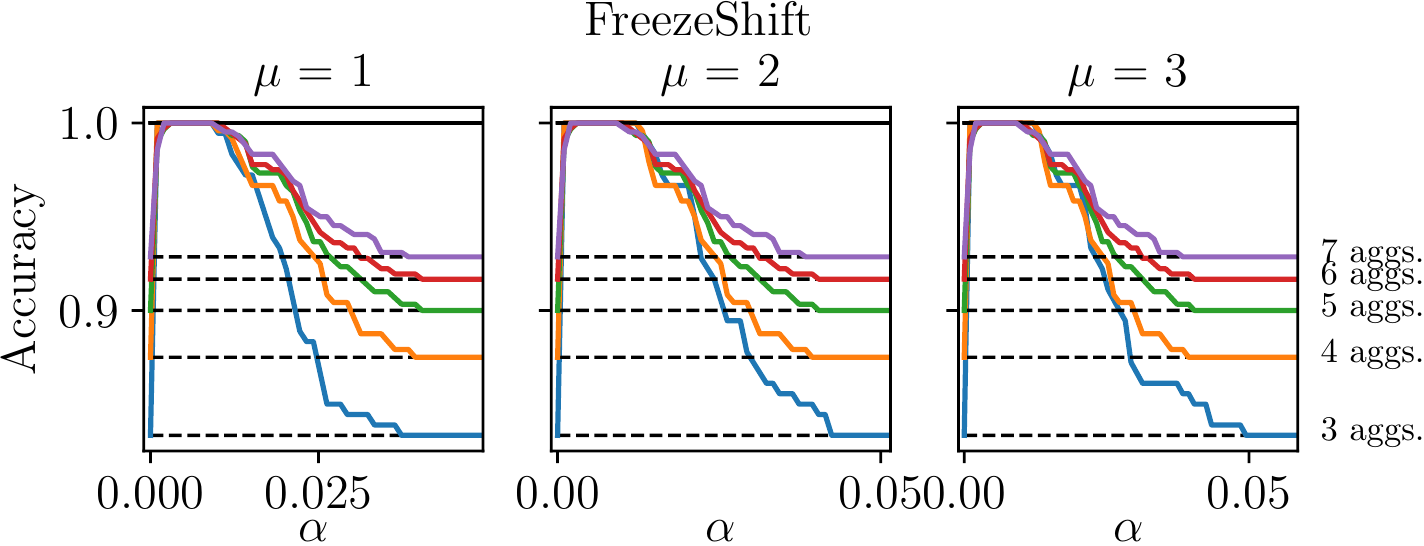}
\includegraphics[width=.9\linewidth]{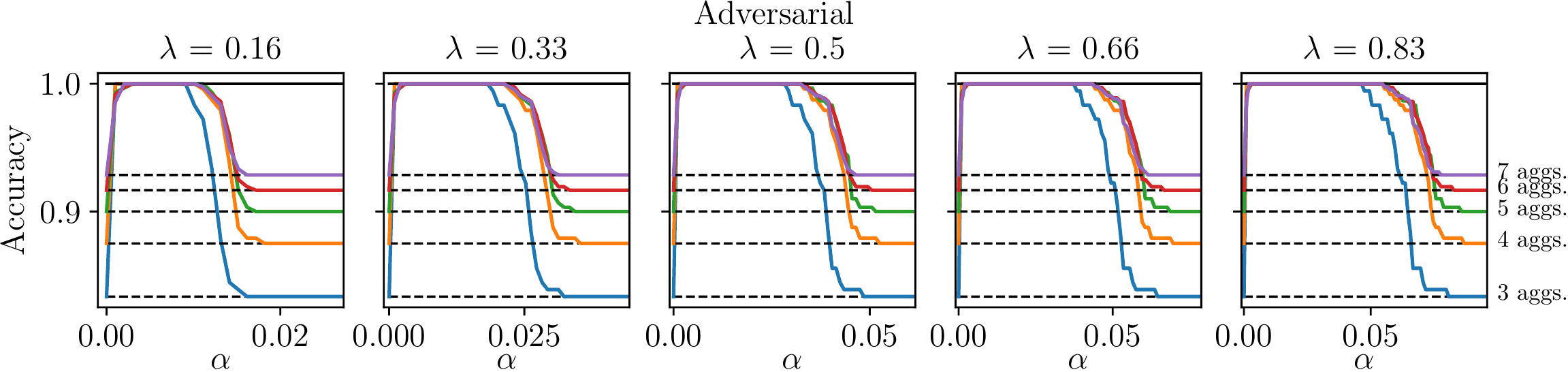}
\caption{Accuracy analyses for each of the proposed attack vectors, attack strengths and number of aggregators. Results averaged over every day of November 2016. All aggregators have capacity for $150\;000$ EVs. Dashed lines represent the naive benchmark for each scenario, which considers every aggregator to be benign.}
\label{fig:accuracyAnalysis1}
\end{figure}

\begin{figure}[htb]
\centering
\includegraphics[width=.9\linewidth]{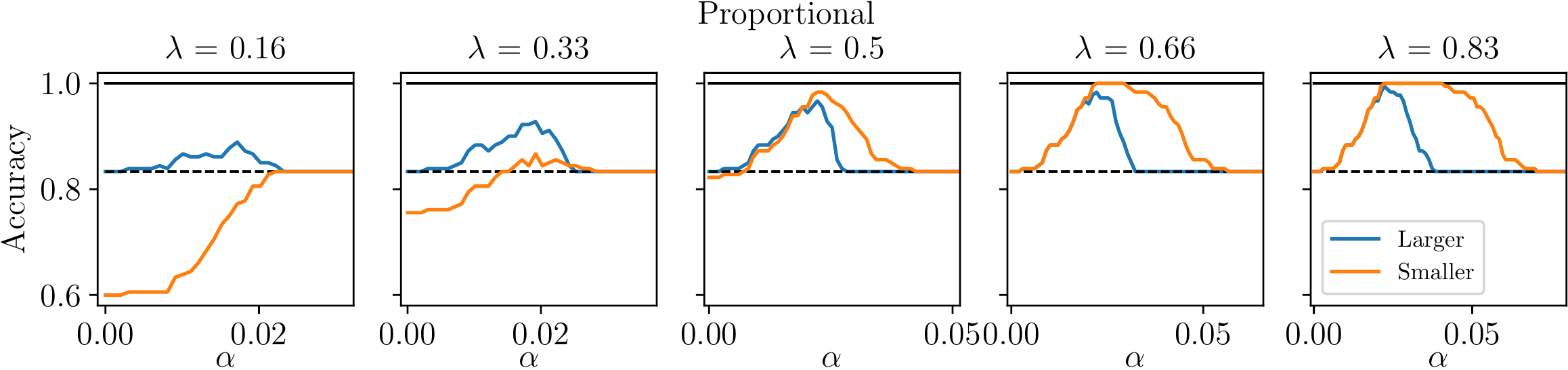}
\includegraphics[width=.58\linewidth]{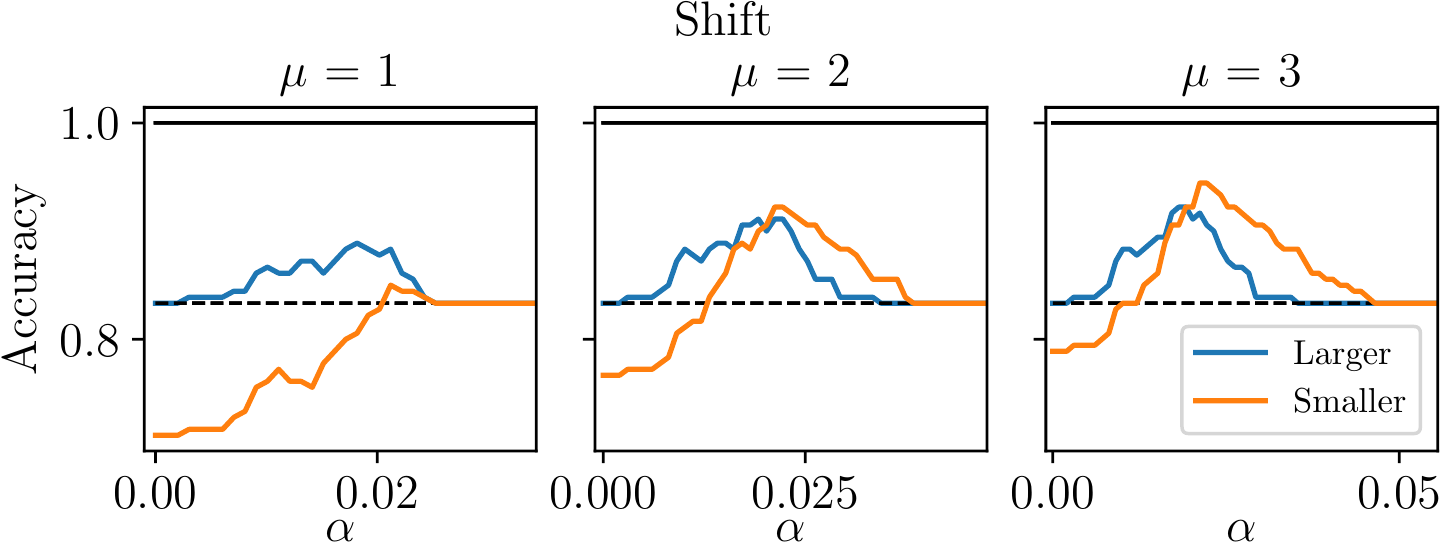}
\includegraphics[width=.9\linewidth]{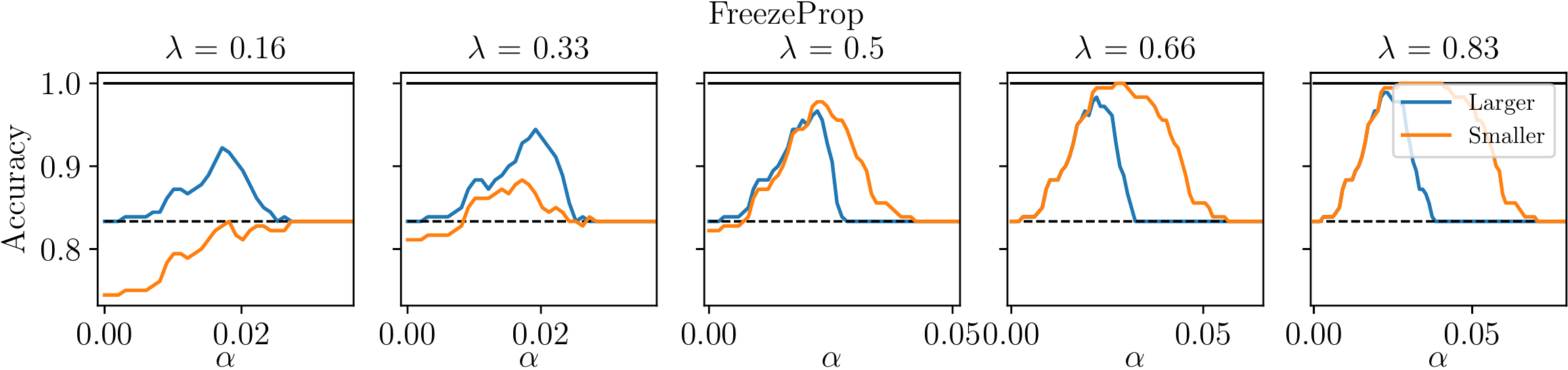}
\includegraphics[width=.58\linewidth]{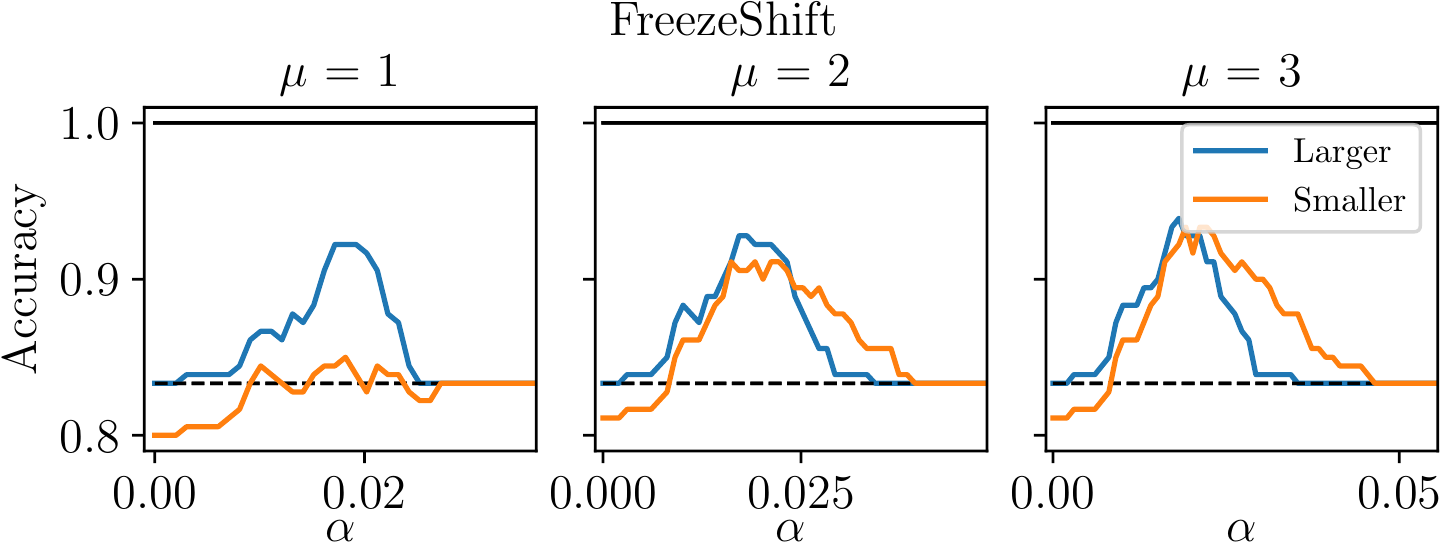}
\includegraphics[width=.9\linewidth]{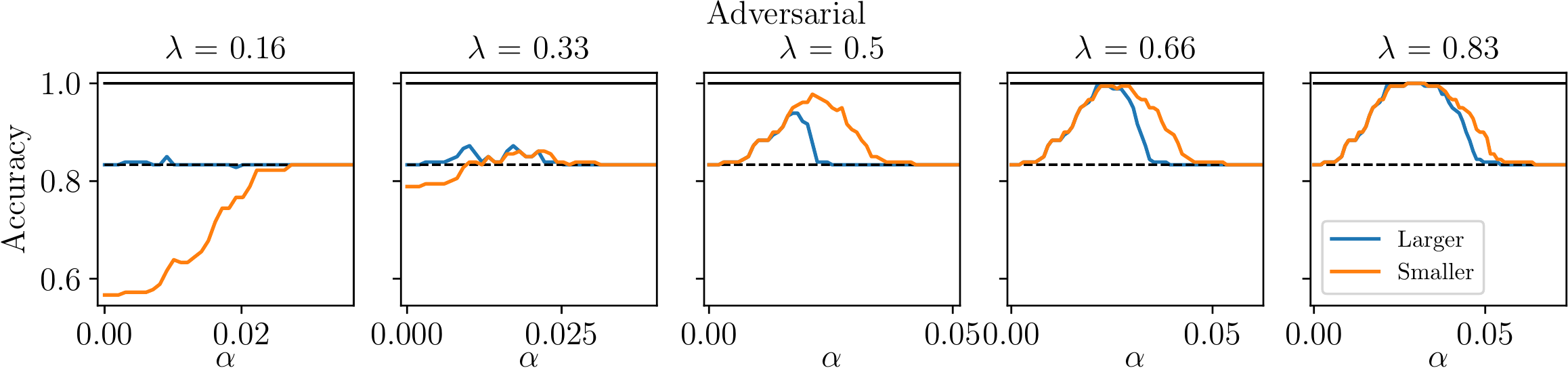}
\caption{Accuracy analyses for each of the proposed attack vectors, attack strengths and aggregator size. Results averaged over every day of November 2016. Dashed lines represent the naive benchmark for each scenario, which considers every aggregator to be benign.}
\label{fig:accuracyAnalysis2}
\end{figure}

So far, we have shown the efficacy of the different proposed attack vectors. Next, we present the empirical performance of the proposed detection algorithm. As mentioned in Section \ref{sec:detManipulation}, the choice of the threshold parameter $\alpha$ is critical for good performance. Recall that the ultimate aim is to maximise correct classifications, \emph{i.e.} true positives and true negatives. If $\alpha$ is set too high, it will fail to detect deviating behaviour. Conversely, if $\alpha$ is set too low, the detection algorithm would be too sensitive and misclassify benign aggregators as deviators.

In order to quantify the performance of the detector we employ the \emph{accuracy} metric \cite{Metz1978}, given by:
$$\textit{Accuracy} = \dfrac{\textit{True Positives} + \textit{True Negatives}}{\textit{Total population}}$$
Specifically, accuracies $0$ and $1$ correspond to detectors which are always wrong and always right, respectively. Moreover, we will compare the accuracy of our proposed algorithm with that of a naive detector that classifies all the aggregators as benign. This is motivated by the fact that our dataset is unbalanced given that we consider at most one deviator per simulation (see Section \ref{sec:detectManipulation}). A well performing algorithm should present increased accuracy from this naive benchmark.

In more detail, we present the results from two different experiments. First, we consider scenarios with varying numbers of aggregators of the same size. This smoothens the natural discrepancies in the difference matrix, $d$, and allows us to focus on each of the different attacks and attack strengths. Second, we present an experiment considering aggregators of different sizes. In this case, the natural discrepancies arising from the size difference play an important role (see Section \ref{sec:quantManipulation}), and it is more difficult for a detection algorithm to distinguish between size effects and manipulation. For each scenario, we simulate every day of November 2016 and use $\rho=10^{-5}$. Recall that we assume that the deviating aggregator performs uses the same attack vector and attack strength in all rounds, and focus our detection algorithm in the first and second rounds. Also, note that we are not interested in the issue of convergence (which is studied in Section \ref{sec:cs_attacks}) and we just explore the performance of the detection algorithm, independent of whether the attacker is being successful or not. The results for each of the two experiments are shown in Figs. \ref{fig:accuracyAnalysis1} and \ref{fig:accuracyAnalysis2}, respectively. Note that the results for each algorithm and its \emph{all} counterpart are very similar and the latter have been omitted.

The results from the first experiment indicate that our proposed algorithm is able to detect the majority of attacks with an accuracy of, or very close to, 1. This shows that our algorithm significantly outperforms the naive benchmark. Moreover, these results are consistent across different numbers of aggregators ranging from 3 to 7. Focusing on each different attack, detection is easier as the strength increases. This is intuitive as stronger attacks have a more pronounced effect on the algorithm and stand out. Interestingly, the attack vector which is most difficult to detect is \emph{Shift}, which is the only one where the algorithm is not able to reach perfect accuracy for any number of aggregators. Finally, the presence of greater number of aggregators makes detection slightly more difficult in most cases. Note that, as explained above, these simulations consider aggregators with the same size. As we will see next, size effects make detection more challenging. Finally, note that for large values of $\alpha$, the detection algorithm classifies everyone as benign thus converging to the naive benchmarks.

In the second experiment, we consider two different settings with three aggregators. Firstly, the attacker has capacity for $450\;000$ EVs, whereas the other two aggregators are smaller and have capacity for $150\;000$ EVs. Secondly, the attacker and one of the benign aggregators have capacity for $150\;000$ EVs, while the attacked aggregator has capacity for $450\;000$ EVs. These two scenarios are called \emph{Larger} and \emph{Smaller} in Fig. \ref{fig:accuracyAnalysis2}, respectively. We can see that size effects make detection considerably more challenging. Our proposed algorithm still outperforms the naive detector in the vast majority of cases, and perfect accuracy is still achieved for high attack strength in some cases, such as \emph{Adversarial}, \emph{Proportional} and \emph{FreezeProp}. However, smaller attacks are difficult to detect and in some cases the performance of our algorithm is comparable to the naive detector (see for example \emph{Adversarial} with $\lambda=0.16$ in the \emph{Smaller} scenario. This suggests that the discrepancies in the difference matrix arising naturally due to size differences are not smoothed enough by the proposed normalisation scheme. However, despite this extra challenge, the proposed detection algorithm significantly outperforms the naive benchmark and presents very good detection results across a variety of settings.

%%%%%%%%%%%%%%%%%%%%%%%%%%%%%%%%%%%%%%%%%%%%%%%%%%%%%%%%%%%%%%%%%%%%%%%%
\subsection{Performance of \emph{Adversarial}}
\label{sec:adversarialAnalysis}

\begin{figure}[!t]
    \centering
    \includegraphics[width=.45\textwidth]{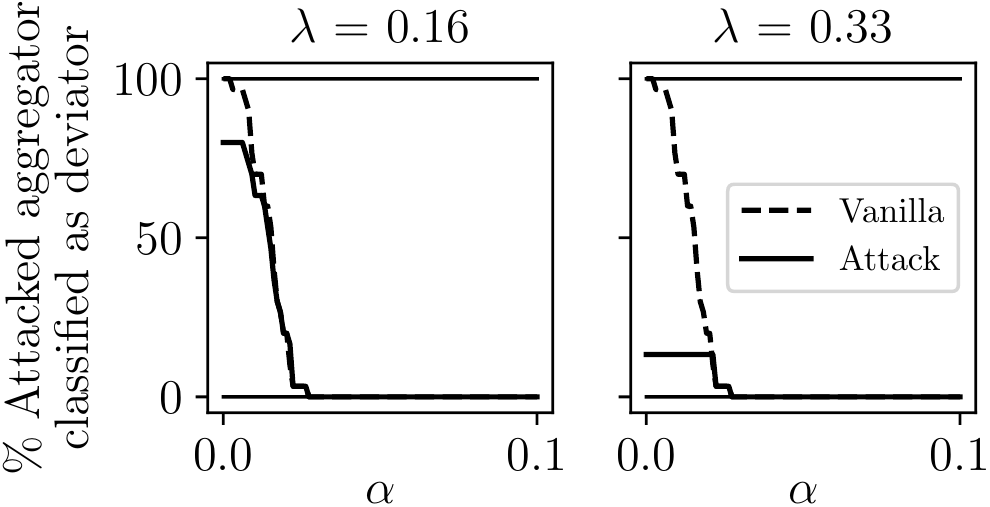}
    \caption{Analysis of the efficacy of \emph{Adversarial} for two attack strengths in the \emph{Smaller} scenario, as described in Section \ref{sec:cs_thresholds}. Dashed lines correspond to the percentage of times the attacked aggregator is incorrectly classified as deviator, in the case with no manipulation. Solid lines represent the same quantity when the aggregator is the target of \emph{Adversarial}. Results averaged over every day of November 2016.}
    \label{fig:adversarialFPs}
\end{figure}

As mentioned in Section \ref{sec:cs_attacks}, the efficacy of the \emph{Adversarial} attack vector is appropriately measured in terms of its success rate in incriminating a benign aggregator as deviator. We perform this analysis using the same datasets as in the previous section: namely the scenarios with varying number of aggregators of the same size and the ones including varying size aggregators. Results indicate that this attack vector is unsuccessful in most cases, not being able to incriminate the attacked aggregator.

The most interesting behaviour is obtained in scenarios containing aggregators of different sizes and where the attacked aggregator has a larger size than the rest, \emph{i.e.} the scenarios labelled as \emph{Smaller} in the previous section (the attacker is \emph{smaller} than the attacked aggregator). In these, the natural discrepancies in the difference matrix cause the detection algorithm to incorrectly classify the attacked aggregator as deviator for small $\alpha$ values. This happens as, for small $\alpha$'s, the algorithm will pick up the aggregator that stands out the most in the difference matrix, even if only by a very small margin. Interestingly, if the attacker performs an \emph{Adversarial} attack on the large benign aggregator, the detection algorithm is able to detect it in some cases, hence eliminating the false positive that would be obtained without manipulation. This situation is depicted in Fig. \ref{fig:adversarialFPs}. In more detail, we can see that weak \emph{Adversarial} attacks ($\lambda=0.16$) achieve up to $80\%$ success rates. However, the same scenario without manipulation presents higher false positive rates due to the reasons explained above. Moreover, increasing the values of $\alpha$ reduces the false positives to zero. Similar results can be observed for stronger attacks ($\lambda=0.33$. For the rest of scenarios, the attacker is not able to incriminate the attacked aggregator at all, and is in turn detected as deviator.

Importantly, we would like to remark the fact that this paper focuses on scenarios where there is, at most, one deviating aggregator (see \ref{sec:quantManipulation}). In more general settings, which will be subject of study in future work, it is likely that this attack vector will present much higher success rates. In more detail, the proposed detection algorithm will choose at most one aggregator as deviator, so the natural deviations arising from size differences compete with the manipulation effects. If size difference stands out more, the detection algorithm will incur a false positive without any manipulation. Conversely, if the effects of an \emph{Adversarial} attack overcome these natural discrepancies, the algorithm will correctly detect the attacker. In the more general case where there are any number of deviating aggregators, and with an algorithm (extending the one proposed in this paper or otherwise) that can detect any number of aggregators as deviators, both size effects and malignant incrimination from an \emph{Adversarial} attack could lead to multiple detections.

%%%%%%%%%%%%%%%%%%%%%%%%%%%%%%%%%%%%%%%%%%%%%%%%%%%%%%%%%%%%%%%%%%%%%%%
%%%%%%%%%%%%%%%%%%%%%%%%%%%%%%%%%%%%%%%%%%%%%%%%%%%%%%%%%%%%%%%%%%%%%%%
\section{Conclusion}
\label{sec:conclusion}

In this paper, we present a decentralised coordination mechanism for multi-EV aggregator bidding in the day-ahead market which employs the Alternating Direction Method of Multipliers (ADMM) algorithm. This proposed algorithm extends previous work in the literature, which addresses the same scenario, but with a centralised framework. Specifically, the proposed decentralised framework removes the need for the aggregators to communicate private requirement information to the coordinator, as each aggregator solves its own local private optimisation problem with their own requirements. This is a key feature for the practical applicability of the proposed coordination mechanism, as real businesses or public service providers would be reluctant to disclose this private information.

Also, we present the first study about strategic manipulation of ADMM algorithms by self-interested internal agents. Even though ADMM and related decentralised optimisation algorithms are widely applied in many disciplines, little work has focused on studying how these algorithms can be disrupted by internal attackers. In order to address this issue, we study how a deviating agent can alter their local algorithm in order to increase their own utility at the expense of their competitors. Focusing on our setting and on our proposed algorithm, we introduce several attack vectors that a self-interested aggregator can employ in order to alter the outcome of the ADMM algorithm. Moreover, in order to prevent strategic manipulation, and working towards resilient decentralised optimisation, we study how deviating behaviour can be detected. In more detail, we propose a mathematical framework which measures the effects that different agents exert onto each other when employing the ADMM algorithm. Furthermore, we propose a threshold-based algorithm which employs this formalism in order to classify the participating aggregators as benign or deviators. Although we focus on an energy setting, the proposed detection framework is general and can be applied in general.

In order to study the proposed decentralised algorithm and detection mechanism, we present an empirical evaluation using real market and vehicle usage data from Spain. We first show the convergence of the decentralised method to the optimal solution for two scenarios, with two and ten cooperating EV aggregators respectively. Convergence can be achieved in around 50 iterations in the first case and around 80 in the second case. Therefore, although problem complexity increases with the number of participants, these numbers suggest the applicability of the algorithm in large settings. With respect to the proposed attack vectors, we analyse their impact on attacker utility (reduced energy costs) and on algorithm convergence. Specifically, we show that an attacker is able to effectively alter the outcome of the algorithm for their own benefit. Finally, we turn our attention to the proposed detection framework, and present an accuracy study to assess its performance. Results show that our algorithm achieves very high accuracy, close and up to 1 in all cases when the considered aggregators have the same size, significantly outperforming the naive benchmark. However, considering aggregators with different sizes makes detection more difficult. Although our proposed algorithm outperforms the naive benchmark in most cases, some small attacks are very challenging to detect.

Several aspects are left for future work. Firstly, the proposed attack vectors in this work are somewhat myopic and do not exploit all the information available to the attacker. More specifically, the attacker could monitor convergence in successive rounds and adjust their strategies accordingly, or devise more complex algorithms that are not simply repeated every round. More sophisticated attack vectors should translate into increased utility gains, better convergence and more difficult detection. Secondly, the proposed detection algorithm can be extended and improved. Right now, it considers each round separately and directly classifies each aggregator as benign or deviator, without taking into account any information about successive rounds. As an example, a more sophisticated model can monitor successive rounds and build a confidence score about an aggregator being a deviator. Thirdly, the same scenario could be studied from a distributed mechanism design perspective (see Section \ref{sec:intro}). In more detail, in order to guarantee strategyproofness, the coordination mechanism needs to truthfully elicit the aggregators requirements and ensure faithful computation. This could be achieved by combining a detection algorithm with appropriate penalties (computation) with the application of an appropriate payment mechanism (elicitation). Finally, it would be interesting to expand the tools proposed in this work to generic multi-agent system ADMM scenarios.

%%%%%%%%%%%%%%%%%%%%%%%%%%%%%%%%%%%%%%%%%%%%%%%%%%%%%%%%%%%%%%%%%%%%%%%
%%%%%%%%%%%%%%%%%%%%%%%%%%%%%%%%%%%%%%%%%%%%%%%%%%%%%%%%%%%%%%%%%%%%%%%
\section*{Acknowledgements}
This work was supported by an EPSRC (UK) Doctoral Training Centre grant (EP/L015382/1). All the data generated and discussed in this work is publicly available \cite{Perez-Diaz2019a}.

\bibliographystyle{theapa}
\bibliography{bibliography}

\end{document}